\documentclass{ws-mpla}
\usepackage[super]{cite}
\usepackage{graphicx}
\usepackage{hyperref}
\usepackage{amsmath, amssymb}
\usepackage{babel}
\usepackage{color}
\usepackage{mathrsfs}
\usepackage{slashed}

\begin{document}

\markboth{Chien-Yeah Seng}
{Towards a discovery of BSM physics from the Cabibbo angle anomaly}

%%%%%%%%%%%%%%%%%%%%% Publisher's Area please ignore %%%%%%%%%%%%%%
\catchline{}{}{}{}{}
%%%%%%%%%%%%%%%%%%%%%%%%%%%%%%%%%%%%%%%%%%%%%%%%%%%%%%%%%%%%%%%%%%%

\title{Towards a discovery of BSM physics from the Cabibbo angle anomaly
}

\author{\footnotesize Chien-Yeah Seng}

\address{Helmholtz-Institut f\"{u}r Strahlen- und Kernphysik and Bethe Center for Theoretical Physics,\\
Universit\"{a}t Bonn, 53115 Bonn, Germany\\
cseng@hiskp.uni-bonn.de}

\maketitle

%\pub{Received (Day Month Year)}{Revised (Day Month Year)}

\begin{abstract}

New developments in both the theories and experiments related to the extraction of the top-row Cabibbo-Kobayashi-Maskawa matrix elements $V_{ud}$ and $V_{us}$ led to a series of new anomalies, for instance the apparent violation of the top-row unitarity relation. It is important to further reduce all the associated Standard Model theory uncertainties in order to better understand whether such observations point towards the possibility of physics beyond the Standard Model, or rather some unexpectedly large Standard Model effects. This requires improved studies of tree-level and higher-order Standard Model corrections that enter the beta decays of pions, neutron, nuclei and kaons. We will briefly review the recent progress along this direction and discuss possible improvements in the future.

%\keywords{Keyword1; keyword2; keyword3.}
\end{abstract}

%\ccode{PACS Nos.: include PACS Nos.}

\section{Introduction}	

This 
%useless
brief review article is prepared according to a remote talk presented by the author in the NT/RIKEN seminar at BNL on 16 October, 2020~\cite{RikenBNL}, but with some information updated to cover the recent developments that occurred after the talk. It is impossible to cover all the important aspects given the limited time of the original seminar, but we hope the readers are still able to get a flavor of the interesting physics under this general topic.   

The Standard Model (SM) of particle physics is arguably one of the most successful theories in physics. However, it fails to explain several important observations in cosmology, e.g. dark energy~\cite{Aghanim:2018eyx,Riess:1998cb,Perlmutter:1998np}, dark matter~\cite{Aghanim:2018eyx,Simon:2019nxf,Salucci:2018hqu,Allen:2011zs} and matter-antimatter asymmetry~\cite{Aghanim:2018eyx,Sakharov:1967dj,Mossa:2020gjc}. At present, all searches of physics beyond the Standard Model (BSM) at high-energy colliders have returned null results; on the other hand, interesting hints have emerged at the ``precision frontier'', which consists of low-energy experiments that measure physical observables to very high precision and look for possible deviations from SM predictions that can be interpreted as signals of BSM physics. In the recent years, we observe a number of anomalies in precision experiments, e.g. the muon $g-2$ anomaly~\cite{Fermigm2,Aoyama:2020ynm,Miller:2007kk,Miller:2012opa,Jegerlehner:2009ry} and the $B$-decay anomalies~\cite{Aaij:2019wad,Aaij:2014ora,Aaij:2015yra,Aaij:2015oid}, which are now collectively known as the flavor anomalies. In this article we focus on a new type of anomaly at the precision frontier which resides in the measured values of the Cabibbo-Kobayashi-Maskawa (CKM) matrix elements from beta decay experiments.

\begin{table}[h]
	\tbl{\label{tab:VudVusnum}Most recent determinations of $|V_{ud}|$, $|V_{us}|$ and $|V_{us}/V_{ud}|$. Lattice results of form factors and decay constants at $N_f=2+1+1$ are adopted in the kaon sector.}
	{\begin{tabular}{|c|c|}
			\hline 
			& $|V_{ud}|$\tabularnewline
			\hline 
			\hline 
			superallowed & 0.97373(31)~\cite{Hardy:2020qwl}\tabularnewline
			\hline 
			$n$ & 0.97377(90)~\cite{Zyla:2020zbs}\tabularnewline
			\hline 
			nuclear mirror & 0.9739(10)~\cite{Hayen:2020cxh}\tabularnewline
			\hline 
			$\pi_{e3}$ & 0.9740(28)~\cite{Feng:2020zdc}\tabularnewline
			\hline 
		\end{tabular}
		\begin{tabular}{|c|c|}
			\hline 
			& $|V_{us}|$\tabularnewline
			\hline 
			\hline 
			$K_{\ell3}$ & 0.22309(56)~\cite{Seng:2021nar}\tabularnewline
			\hline 
			$\tau$ & 0.2221(13)~\cite{HFLAV:2019otj}\tabularnewline
			\hline 
			Hyperon & $0.2250(27)$~\cite{Cabibbo:2003ea}\tabularnewline
			\hline 
		\end{tabular}
		\begin{tabular}{|c|c|}
			\hline 
			& $|V_{us}/V_{ud}|$\tabularnewline
			\hline 
			\hline 
			$K_{\mu2}/\pi_{\mu2}$ & 0.23131(51)~\cite{Seng:2021nar}\tabularnewline
			\hline 
			$K_{\ell3}/\pi_{e3}$ & 0.22908(87)~\cite{Seng:2021nar}\tabularnewline
			\hline 
		\end{tabular} }
	\end{table}

Beta decays of hadrons and nuclei provided some of the most important experimental foundations for the understanding of the charged weak interaction in SM, including the neutrino postulation by Pauli~\cite{Pauli:1949zm}, the confirmation of parity (P)-violation~\cite{Wu:1957my,Lee:1956qn}, the proposal of the V-A structure in the charged weak sector~\cite{Feynman:1958ty,Sudarshan:1958vf}, and the discovery of the CKM matrix
\begin{equation}
V_{\mathrm{CKM}}=\left(\begin{array}{ccc}
V_{ud} & V_{us} & V_{ub}\\
V_{cd} & V_{cs} & V_{cb}\\
V_{td} & V_{ts} & V_{tb}
\end{array}\right)
\end{equation}
that mixes the quark flavor eigenstates to form mass eigenstates~\cite{Cabibbo:1963yz,Kobayashi:1973fv}. In the modern days, beta decays provide stringent tests of SM; in particular, combining precise measurements of decay rates/branching ratios and appropriate SM theory inputs, one could determine the top-row CKM matrix elements $V_{ud}$ and $V_{us}$ to high precision, which are then used to check against SM predictions. In Table~\ref{tab:VudVusnum} we summarize the most recent determinations of $V_{ud}$ and $V_{us}$ from different charged weak decay processes.

\begin{figure}[ph]
	\centerline{\includegraphics[width=3.0in]{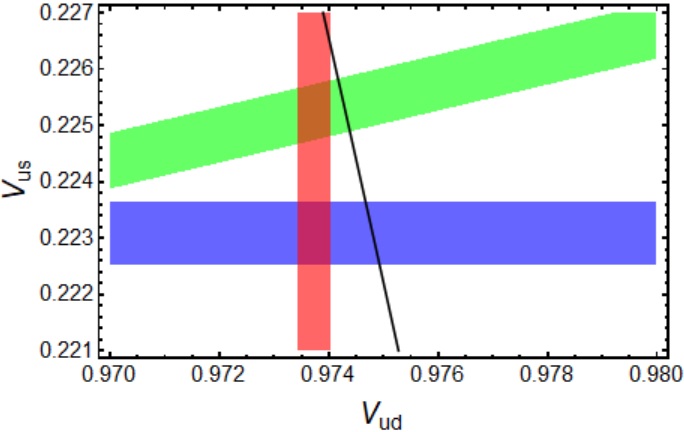}}
	\vspace*{8pt}
	\caption{A combined plot of $|V_{ud}|$ from superallowed decays (red band), $|V_{us}|$ from $K_{\ell 3}$ (blue band), $|V_{us}/V_{ud}|$ from $K_{\mu 2}/\pi_{\mu 2}$ (green band) and the SM unitarity requirement (black line).\protect\label{fig:VudVus}}
\end{figure}

 An important consequence of $V_\mathrm{CKM}$ being a unitary matrix is the following ``top-row CKM unitarity relation'':
\begin{equation}
\Delta_\mathrm{CKM}^u\equiv|V_{ud}|^2+|V_{us}|^2+|V_{ub}|^2-1=0~.\label{eq:unitarity}
\end{equation}
In practice, since $|V_{ub}|^2\sim 10^{-5}$ is much smaller than even the quoted uncertainties in $|V_{ud}|^2$ and $|V_{us}|^2$, we can drop this matrix element in our analysis, and thus Eq.\eqref{eq:unitarity} reduces to the simpler Cabibbo unitarity relation: $|V_{ud}|^2+|V_{us}|^2=1$, which allows us to parameterize the two matrix elements with a single Cabibbo angle $\theta_C$: $|V_{ud}|=\cos\theta_C$, $|V_{us}|=\sin\theta_C$. However, taking the 
latest independent determinations of these matrix elements, we find that such relation is not quite satisfied. For instance, the most recent online version of the Particle Data Group (PDG) review quoted the following unitary sum~\cite{Zyla:2020zbs}:
\begin{equation}
\Delta_\mathrm{CKM}^u=-0.0015(6)_{V_{ud}}(4)_{V_{us}}~,
\end{equation} 
with $|V_{ud}|$ taken from superallowed decays and $|V_{us}|$ taken from the average of the semileptonic ($K_{\ell 3}$) and leptonic ($K_{\mu 2}$) kaon decay results (and also average over the results with $N_f=2+1$ and $N_f=2+1+1$ lattice inputs). So from the above we observe an apparent violation of unitarity at the level of $2\sigma$. This is, however, not the whole story because we also observe from Table~\ref{tab:VudVusnum} a $\sim 2.8\sigma$ disagreement between the values of $|V_{us}|$ extracted from $K_{\ell 3}$ and $K_{\mu 2}$ as we will discuss in more detail later. Therefore, different choices of $|V_{us}|$ may lead to different degrees of unitarity violation, for example $|V_{ud}|$ from superallowed beta decays and $|V_{us}|$ from $K_{\ell 3}$ combine to give:
\begin{equation}
|V_{ud}|_{0^+}^2+|V_{us}|_{K_{\ell 3}}^2-1=-0.0021(7)~,
\end{equation}
i.e. a $\sim 3\sigma$ deviation from unitarity. This series of anomalies imply the mutual disagreement in the measurement of the Cabibbo angle $\theta_C$ from different beta decay experiments (see Fig.\ref{fig:VudVus}, where the implied Cabibbo angle from each colored band is deduced from its overlap with the black line), and therefore are also known collectively as the Cabibbo angle anomaly. They have attracted world-wide attentions in the recent years concerning possible implications on BSM physics~\cite{Crivellin:2020lzu,Crivellin:2021njn}, featured in international conferences and workshops \cite{UMass18,TAMU19,ECT19,UMass19,INT19,APS20,MITP20,Snowmass20,Moriond21,DPG21,DNP21,CKM21} and at least 6 letters of interest in SnowMass 2021 \cite{BaccaSnowmass,BaesslerSnowmass,ArevaloSnowmass,BazavovSnowmass,BoyleSnowmass,BhattacharyaSnowmass}. In order to better appreciate the discovery potential of BSM physics from these anomalies, we shall briefly review the recent progress in the extraction of $|V_{ud}|$ and $|V_{us}|$ from various beta decays which led to the current situation, and discuss the desired improvements in the future.

\section{$|V_{ud}|$}

Following the original seminar, we discuss the extraction of $|V_{ud}|$ from three types of beta decays, namely the pion semileptonic beta decay ($\pi_{e3}$), the free neutron decay and superallowed nuclear decays.

\subsection*{$\pi_{e3}$} 
$|V_{ud}|$ is extracted from the semileptonic pion decay process $\pi^+\rightarrow\pi^0 e^+\nu_e(\gamma)$ through the following master formula:
\begin{equation}
	\Gamma_{\pi_{e3}}=\frac{G_F^2|V_{ud}|^2M_{\pi^+}^5|f_+^\pi(0)|^2}{64\pi^3}(1+\delta_\pi)I_\pi~,
\end{equation}
where $\Gamma_{\pi_{e3}}$ is the partial decay width, $G_F=1.1663787(6)\times 10^{-5}$~GeV$^{-2}$ is the Fermi's constant measured from muon decay~\cite{MuLan:2012sih}, $f_+^\pi(0)$ is the $\pi^+\rightarrow\pi^0$ transition form factor at zero momentum transfer, $\delta_\pi$ is the electroweak radiative correction (RC), and $I_\pi$ is a known phase space factor.

$\pi_{e3}$ is an extremely clean channel from the theory perspective. First of all, since pions are spinless, the tree-level transition form factor probes only the conserved vector current (CVC), of which matrix element is simply fixed by isospin symmetry: $|f_+^\pi(0)|=1$. Isospin symmetry breaking (ISB) corrections originate from the quark mass splitting scales as $\mathcal{O}((m_d-m_u)^2)$ instead of $\mathcal{O}(m_d-m_u)$ due to the Behrends-Sirlin-Ademollo-Gatto theorem~\cite{Behrends:1960nf,Ademollo:1964sr}, and is numerically insignificant. Also, given the small $\pi^+-\pi^0$ mass splitting, the $t$-dependence of the form factor is negligible as well. Therefore, to our desired precision level the electroweak RC factor $\delta_\pi$ is the only non-trivial theory input. Early analysis by Sirlin~\cite{Sirlin:1977sv} and later studies based on Chiral Perturbation Theory (ChPT)~\cite{Cirigliano:2002ng,Cirigliano:2003yr} all carry an irreducible theory uncertainty of the generic size $10^{-3}$, which reflects the incalculable effects of non-perturbative Quantum Chromodynamics (QCD) at the hadronic scale. This barrier is, however, overcome recently by a new lattice QCD calculation which we will discuss more later~\cite{Feng:2020zdc}. This new input fixes $\delta_\pi$ to an unprecedented precision level of $10^{-4}$, and makes $\pi_{e3}$ the theoretically cleanest avenue to extract $|V_{ud}|$. 

The major limiting factor in $\pi_{e3}$ is the experimental uncertainty the branching ratio $\mathrm{BR}(\pi_{e3})$. The small central value $\sim 10^{-8}$ makes its precise determination very challenging. Currently the best measurement of this quantity comes from PIBETA experiment published in 2004~\cite{Pocanic:2003pf}:
\begin{equation}
\mathrm{BR}(\pi_{e3})=1.038(6)\times 10^{-8}~.
\end{equation}
Notice that the value quoted above is slightly larger than that in the original paper to account for the effect of the updated $\mathrm{BR}(\pi_{e2})$ normalization~\cite{Czarnecki:2019iwz}. Partially motivated by the recent improvements from the theory side, a next-generation rare pion decay experiment known as PIONEER is proposed and aims to improve the experimental precision of $\mathrm{BR}(\pi_{e3})$ first by a factor of 3 ~\cite{Pioneer,ArevaloSnowmass}, and later by a factor 10 as a long-term goal.

\subsection*{Free neutron decay}

The decay of free neutron is a mixed transition that involves not just the Fermi, but also the Gamow-Teller (GT) matrix element at tree level; the latter probes the non-conserved axial current and must be determined through separate measurements. The master formula for the $|V_{ud}|$ extraction reads~\cite{Zyla:2020zbs}:
\begin{equation}
|V_{ud}|^2=\frac{5024.7~\mathrm{s}}{\tau_n(1+3\lambda^2)(1+\Delta_R^V)}~,\label{eq:masterneutron}
\end{equation}
where $\tau_n$ is the neutron lifetime, $\lambda=g_A/g_V$ is the (renormalized) ratio between the neutron axial and vector coupling constant, and $\Delta_R^V$ is the so-called ``single-nucleon inner RC'' which is the only non-trivial theory input and the major source of theory uncertainty. We will discuss this quantity in more detail in the later subsection, but here let us just quote one of the several most recently-adopted values: $\Delta_R^V=(2.454\pm 0.019)$\%~\cite{Hardy:2020qwl}. 

Similar to $\pi_{e3}$, currently the major limiting factor in the $|V_{ud}|$ extraction from neutron decay comes from experiment, namely the measurements of $\tau_n$ and $\lambda$. For the neutron lifetime, although the recent UCN$\tau$ experiment at Los Alamos has achieved a measurement with an unprecedented level of precision: $\tau_n=877.75\pm 0.28_\mathrm{stat}+0.22/-0.16_\mathrm{syst}~\mathrm{s}$~\cite{UCNt:2021pcg}, the long-standing discrepancy between the lifetime measurement with the ``beam'' and ``bottle'' methods, where the latter generically reports a $\sim 10$~s longer lifetime, still persists and requires a resolution~\cite{Fornal:2018eol}\footnote{The results from the ``beam'' measurements are currently not included in the PDG average.}. In the meantime, the ratio $\lambda$ is measured through various correlation coefficients in the free neutron differential decay rate:
\begin{equation}
d\Gamma\propto 1+a\frac{\vec{p}_e\cdot\vec{p}_\nu}{E_eE_\nu}+b\frac{m_e}{E_e}+\hat{n}\cdot\left[A\frac{\vec{p}_e}{E_e}+B\frac{\vec{p}_\nu}{E_\nu}+...\right]
\end{equation}
with $\hat{n}$ the neutron polarization.
Currently, the single most precise determination of $\lambda$ comes from the PERKEO III experiment through the measurement of the asymmetry parameter $A$: $\lambda=-1.27641(45)_\mathrm{stat}(33)_\mathrm{syst}$~\cite{Markisch:2018ndu}. However, in PDG a scale factor $S=2.7$ is included to account for an observed discrepancy between the values of $\lambda$ deduced from the parameters $A$ and $a$ respectively; the latest input to the latter comes from the aSPECT experiment, which returned $\lambda=-1.2677(28)$~\cite{Beck:2019xye}. This
leads to a larger quoted uncertainty in the PDG average of $\lambda$~\cite{Zyla:2020zbs}. In summary, the measurements of $\tau_n$ and $\lambda$ must reach a relative precision of $0.019$\% and $0.011$\% respectively in order to be compatible with the precision level of the theory input for the $|V_{ud}|$ extraction.

\subsection*{Superallowed nuclear decays} 

Finally, we discuss the so-called ``superallowed'' nuclear decay, which means the beta decay of a $J^P=0^+$ nucleus to another $J^P=0^+$ nucleus. The measured quantity in this decay is the so-called $ft$-value, where $t$ is the half-life and $f$ is a known statistical function. The master formula for the $|V_{ud}|$ extraction reads~\cite{Hardy:2020qwl}:
\begin{equation}
|V_{ud}|^2=\frac{2984.43~\mathrm{s}}{\mathcal{F}t(1+\Delta_R^V)}~,\label{eq:Master0+}
\end{equation}
where $\Delta_R^V$ is the same single-nucleon inner RC as in Eq.\eqref{eq:masterneutron}, and $\mathcal{F}t$ is the nuclear-structure (NS)-corrected $ft$ value of the decay, which is supposed to be nucleus-independent assuming that SM is the correct underlying theory. The experimental uncertainty in $\mathcal{F}t$ is greatly reduced by averaging over 15 best-measured superallowed transitions, whose $ft$-value precision is 0.23\% or better~\cite{Hardy:2020qwl}. This makes the superallowed nuclear decays currently the best avenue for the $|V_{ud}|$ extraction, and the only channel in which the dominant sources of uncertainties are from theory instead of experiment. We will discuss the complexities associated to the NS corrections to $\mathcal{F}t$ in the later subsection.

In what follows, we discuss the important higher-order SM theory inputs for the $|V_{ud}|$ extraction from the decay processes above. There are in general three types of such corrections, namely (1) recoil corrections, (2) ISB corrections and (3) electroweak RC. The first type is well-studied and is less of an issue~\cite{Holstein:1974zf,Wilkinson:1982hu,Gudkov:2008pf,Ivanov:2012qe,Ivanov:2020ybx}; the second type is suppressed as $\mathcal{O}((m_d-m_u)^2)$ and is negligible in $\pi_{e3}$ and free neutron decay, but is relevant in superallowed decays due to nuclear enhancements and is usually classified as a part of the NS-correction; the third type is non-trivial in all decay processes and will comprise most of our discussions below.

\subsection{Radiative corrections: Overview}

\begin{figure}
	\begin{centering}
		\includegraphics[scale=0.5]{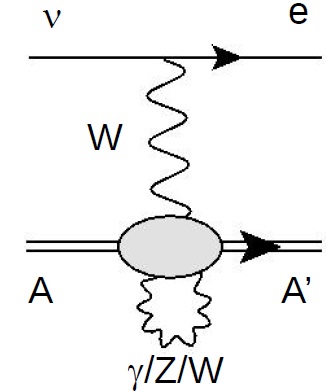}
		\includegraphics[scale=0.5]{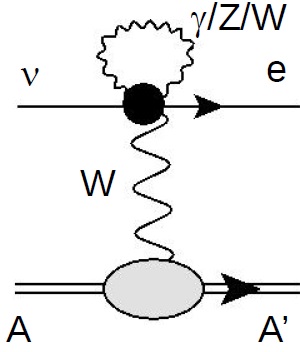}
		\includegraphics[scale=0.5]{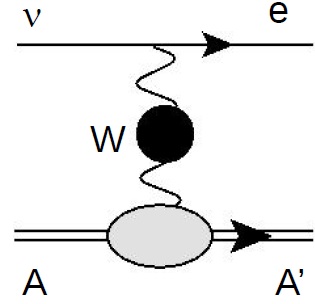}
		\includegraphics[scale=0.5]{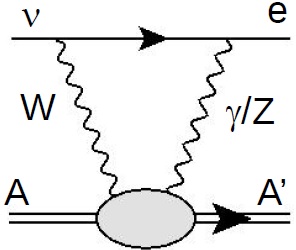}
		\includegraphics[scale=0.5]{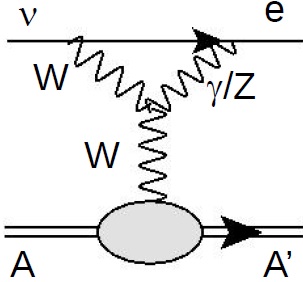}
		\includegraphics[scale=0.5]{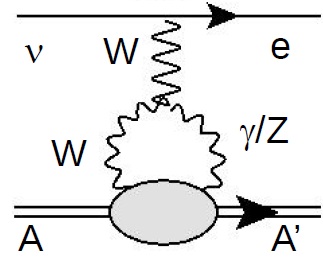}\hfill
		\par\end{centering}
	\caption{\label{fig:EWRC}One-loop electroweak radiative corrections in pion, neutron and nuclear beta decays. Notice that bremsstrahlung (i.e. real photon emission) contributions have to be added to cancel the infrared divergences. }
\end{figure}

\begin{figure}
	\begin{centering}
		\includegraphics[scale=0.5]{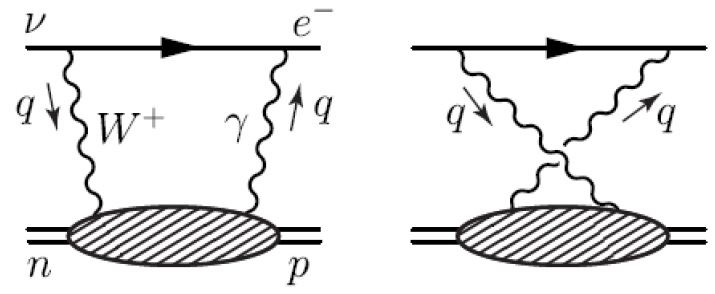}\hfill
		\par\end{centering}
	\caption{\label{fig:gammaW}The free neutron axial $\gamma W$-box diagrams.}
\end{figure}

The generic one-loop electroweak RC to the pion/neutron/superallowed beta decays are summarized in Fig.\ref{fig:EWRC}. They share a similarity, namely the initial and final strongly-interacting systems are almost degenerate. In this special case, the classical current algebra (CA) analysis by Sirlin~\cite{Sirlin:1977sv} shows that most of these RCs are either reabsorbed into the definition of the Fermi's constant $G_F$ or calculable to satisfactory precision ($10^{-5}$) independent of the details of hadronic structures\footnote{See also Ref.\cite{Seng:2021syx} for a detailed review.}. The only exception is the pair of diagrams that involve the simultaneous exchange of a photon and a $W$-boson, the latter couples to the axial charged weak current at the hadron side (see Fig.\ref{fig:gammaW} for the case of free neutron). Bill Marciano often refers them as ``the infamous (axial) $\gamma W$-box diagrams'', because they form the primary source of theory uncertainties in the $|V_{ud}|$ extraction. These diagrams can be expressed as an integral over the loop momentum $q$; at large $Q^2=-q^2$, the integrand is perturbatively calculable, but at small $Q^2$ it is sensitive to the details of hadronic structures, which are governed by the non-perturbative QCD. Also, the transition point between the perturbative and non-perturbative regimes is ambiguous. 

\subsubsection{\label{sec:singlenucleon}Single-nucleon sector}

Decade-long investigations of the RC, in particular the free neutron axial $\gamma W$-box diagrams, had been performed by Marciano and Sirlin. In 1983, they estimated the loop integral by utilizing perturbative QCD (pQCD) at large-$Q^2$ and elastic form factors at small-$Q^2$, with a varying perturbative matching point as an estimate of the theory uncertainty~\cite{Marciano:1982mm,Marciano:1983ss}. Leading two-loops effects were estimated in 2004 together with Czarnecki~\cite{Czarnecki:2004cw}. In a seminal paper in 2006~\cite{Marciano:2005ec}, they pointed out that the pQCD correction to the axial box diagram is identical to that of the polarized Bjorken sum rule~\cite{Bjorken:1966jh,Bjorken:1969mm} in the chiral limit, to all orders in $\alpha_s$. This allowed them to make use of the $\mathcal{O}(\alpha_s^3)$ pQCD expression available at that time~\cite{Larin:1990zw} to improve their large-$Q^2$ prediction. In the meantime, they continued to adopt the elastic form factors at small-$Q^2$, while improving their predictions at intermediate distances by introducing a three-resonance interpolating function. All of these combined to give $\Delta_R^V=0.02361(38)$, which was taken as the state-of-the-art result for more than a decade. Together with all other theory and experimental inputs, PDG 2018 reported $\Delta_\mathrm{CKM}^u=-0.0006(5)$, in good agreement with unitarity~\cite{Tanabashi:2018oca}.

\begin{figure}
	\begin{centering}
		\includegraphics[scale=0.3]{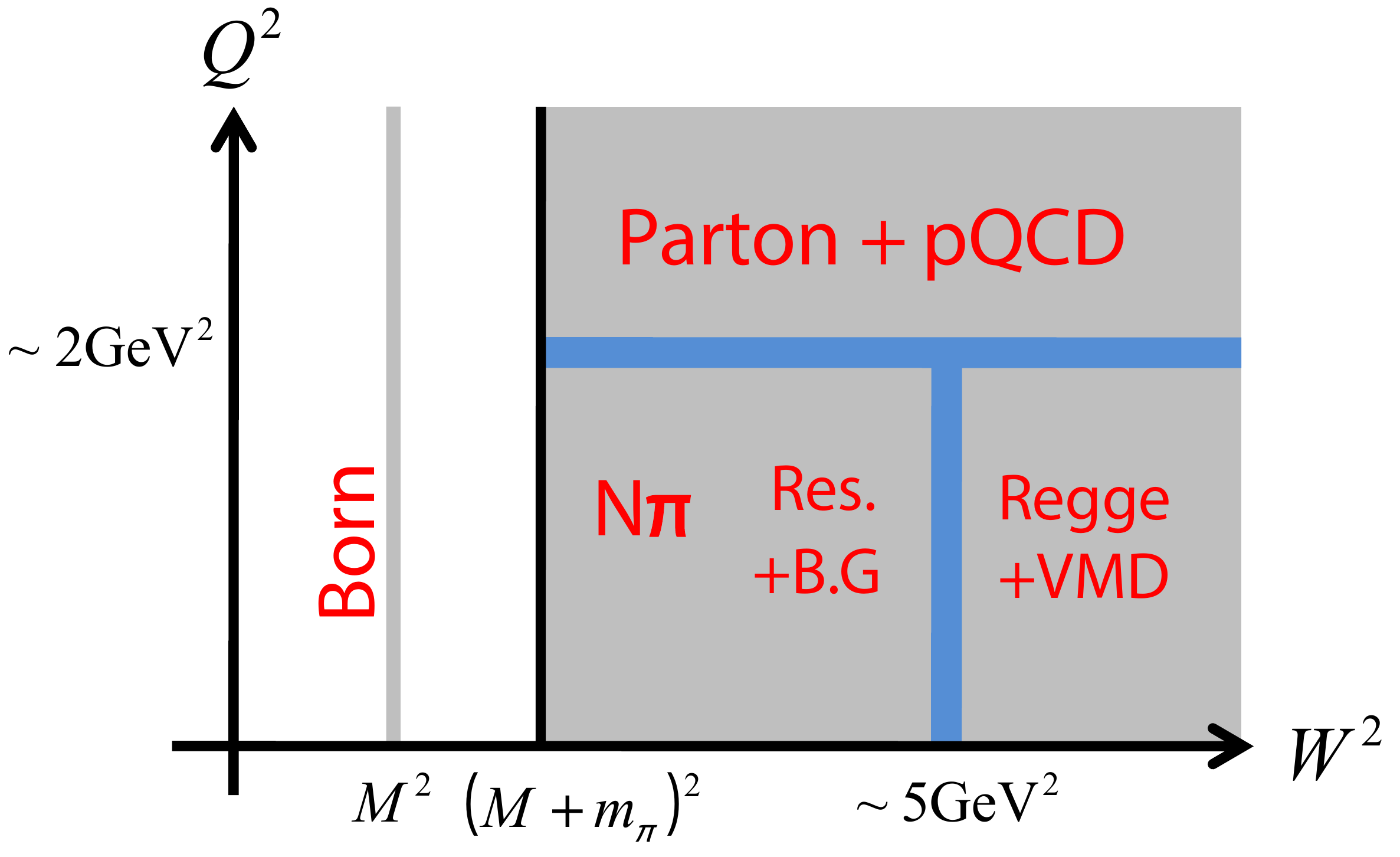}\hfill
		\par\end{centering}
	\caption{\label{fig:WQ2}The ``phase space diagram'' of the structure function $F_3^{(0)}$. }
\end{figure}

A new breakthrough occurred in late 2018, where a novel dispersion relation (DR) treatment was introduced to study the free neutron axial $\gamma W$-box diagram~\cite{Seng:2018yzq,Seng:2018qru}. The key is to rewrite the box diagram contribution as the following integral:
\begin{equation}
\Box_{\gamma W}^{VA}=\frac{3\alpha}{2\pi}\int_0^\infty \frac{dQ^2}{Q^2}\frac{M_W^2}{M_W^2+Q^2}M_3^{(0)}(1,Q^2)~,\label{eq:gammaWbox}
\end{equation}
where
\begin{equation}
M_3^{(0)}(1,Q^2)=\frac{4}{3}\int_0^1 dx\frac{1+2r}{(1+r)^2}F_3^{(0)}(x,Q^2)\end{equation}
(with $x=Q^2/(2p\cdot q)$ the usual Bjorken variable, and $r\equiv \sqrt{1+4m_N^2x^2/Q^2}$) is the so-called ``first Nachtmann moment'' of the spin-independent, parity-odd structure function $F_3^{(0)}(x,Q^2)$~\cite{Nachtmann:1973mr,Nachtmann:1974aj}, the latter is defined through the following forward hadronic tensor:
\begin{eqnarray}
W^{(0)\mu\nu}_{\gamma W}(p,q)&=&\frac{1}{8\pi}\int d^4 z e^{iq\cdot z}\left\langle p(p)\right|[J_\mathrm{em}^{(0)\mu}(z),J_W^\nu(0)]\left|n(p)\right\rangle\nonumber\\
&=&\frac{1}{8\pi}\sum_X (2\pi)^4\delta^{(4)}(p+q-p_X)\left\langle p(p)\right|J_\mathrm{em}^{(0)\mu}\left|X\right\rangle\left\langle X\right|J_{W}^{\nu}\left|n(p)\right\rangle\nonumber\\
&=&\left(-g^{\mu\nu}+\frac{q^\mu q^\nu}{q^2}\right)F_1^{(0)}+\frac{\hat{p}^\mu\hat{p}^\nu}{p\cdot q}F_2^{(0)}-i\varepsilon^{\mu\nu\alpha\beta}\frac{q_\alpha p_\beta}{2p\cdot q}F_3^{(0)}+...
\end{eqnarray}
Here $J_W^\mu$ is the charged weak current, and $J_\mathrm{em}^{(0)\mu}$ is the isosinglet component of the electromagnetic current.
Unlike all the previous treatments that analyze the hadronic function in terms of a single kinematic variable $Q^2$, in the DR representation one is able to access the details of hadron physics in terms of two kinematic variables, namely $Q^2$ and $x$ (or equivalently, $W^2=(p+q)^2$) in order to identify the dominant on-shell intermediate states $X$ that contribute to the structure function $F_3^{(0)}(x,Q^2)$.

Based on general knowledge of hadron physics, one expects the dominant contributors at different regions of $\{W^2,Q^2\}$ to be depicted by the ``phase space diagram'' in Fig~\ref{fig:WQ2}: the single-nucleon elastic (Born) pole occurs at $W^2=m_N^2$, and the inelastic continuum begins at the pion production threshold $W^2=(m_N+M_\pi)^2$. Along the $Q^2$ direction, perturbative QCD starts to work above $Q^2\sim 2$~GeV$^2$ (with evidences coming from experiments and lattice as we will explain later). In general, these contributions fall into two categories:
\begin{enumerate}
	\item The ``non-asymptotic'' pieces (Born, low-energy continuum, resonances): They are very different for different $F_3$, and need to be calculated explicitly case-by-case. Fortunately, the required experimental inputs (e.g. the form factors) are easily available and the uncertainties are under control.
	\item The ``asymptotic'' pieces (perturbative regime, Regge regime): These contributions are largely universal for different $F_3$, up to calculable multiplicative factors. They can either be calculated from first-principles, or inferred from experimental data of other measurable structure functions.
\end{enumerate}

We may now analyze the inputs to the integral in Eq.\eqref{eq:gammaWbox} adopted in Ref.\cite{Seng:2018yzq}. We first discuss the non-asymptotic contributions: The isosinglet magnetic Sachs form factor and the axial form factor that appear in the Born contribution were taken from Refs.\cite{Ye:2017gyb,Bhattacharya:2011ah}. The $N\pi$ contribution was calculated using baryon ChPT at tree level, with extra $Q$-dependence modeled by inserting the electroweak form factors; its contribution is numerically small. For the resonance piece, it turns out that in $F_3^{(0)}$ only the $I=1/2$ resonances contribute assuming isospin symmetry, which excludes the largest $\Delta$-contributions. The resonance parameters were taken from Refs.\cite{Drechsel:2007if,Lalakulich:2006sw}, but their sizes are again very small. 

\begin{figure}
	\begin{centering}
		\includegraphics[scale=1]{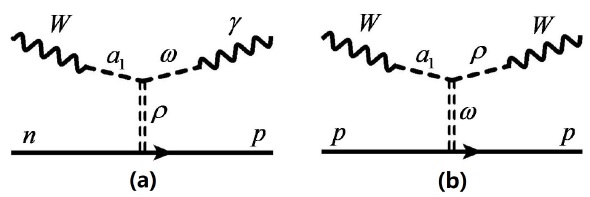}\hfill
		\par\end{centering}
	\caption{\label{fig:Regge}The leading Regge-exchange diagram contributing to  $F_3^{(0)}$ (left) and $F_3^{\nu p+\bar{\nu}p}$ (right) respectively. The broken vertical line represents the exchanged Regge trajectory.}
\end{figure}

Next we turn to the asymptotic contributions. First, at large $Q^2$, one again makes use of the available expression of pQCD corrections to the polarized Bjorken sum rule as pointed out before. The more non-trivial region resides at small $Q^2$ and large $W^2$, from which the multi-hadron intermediate state contributions can be described economically using a Regge-exchange picture depicted in Fig.\ref{fig:Regge}. This allows us to relate the Regge contribution to $F_{3}^{(0)}$ to that of a separate spin-independent, parity-odd structure function $F_{3}^{\nu p+\bar{\nu}p}$ involving the product of two isovector currents; the latter is measurable from neutrino/antineutrino-nucleus scattering experiments~\cite{Kataev:1994rj,Kim:1998kia,Bolognese:1982zd,Allasia:1985hw} after properly subtracting out the non-universal pieces. Due to the near-degeneracy between the $\rho$ and $\omega$ trajectories, the major differences between the two diagrams in Fig.\ref{fig:Regge} come from the gauge boson-vector meson mixing parameters, as well as the coupling constants between the nucleon and the vector mesons. The relations between these parameters of the two cases are predicted by the spin-flavor SU(6) symmetry~\cite{deSwart:1963pdg}. That leads to the following matching:
\begin{equation}
\frac{M_{3,\mathrm{Regge}}^{(0)}(1,Q^2)}{M_{3,\mathrm{Regge}}^{\nu p+\bar{\nu}p}(1,Q^2)}\approx\frac{1}{36}~,
\end{equation}  
i.e. the experimental inputs of $M_{3,\mathrm{Regge}}^{\nu p+\bar{\nu}p}(1,Q^2)$ help us to fix $M_{3,\mathrm{Regge}}^{(0)}(1,Q^2)$. There is a residual model-dependence in the matching relation above, but at this point the major source of uncertainty comes from the neutrino scattering data. 

Collecting all the information above, Refs.\cite{Seng:2018yzq,Seng:2018qru} reported a new value of $\Delta_R^V=0.02467(22)$ which has a better precision than the Marciano-Sirlin result, and at the same time with a significantly shifted central value; this is mainly due to the underestimated intermediate-distance contributions from the latter. The new result led directly to a reduction of the central value of $|V_{ud}|$, which provided the first hint in the recent years of the top-row CKM unitarity deficit. This shift was later confirmed by several independent studies~\cite{Czarnecki:2019mwq,Seng:2020wjq,Hayen:2020cxh,Shiells:2020fqp}.

\subsubsection{Inputs from lattice QCD}

The precision of the dispersive analysis is limited by the quality of the neutrino scattering data, which is particularly poor around $Q^2\sim 1$~GeV$^2$ where the results of the new and old analysis differ by the most\footnote{On the other hand, the more precise data at large $Q^2$ by the CCFR group~\cite{Kataev:1994rj,Kim:1998kia} confirms the validity of the pQCD calculation at $Q^2>2$~GeV$^2$. }. Better-quality data may come from the Deep Underground Neutrino Experiment (DUNE) which is, in any case, not in reach in the near future~\cite{Acciarri:2016crz,Alvarez-Ruso:2017oui}. To overcome such limitations, one may turn to first-principles calculations. Fully-inclusive lattice calculation of the RC (i.e. virtual + real) is not the best choice for pion and neutron beta decays, because as we emphasized before, the only non-trivial piece is really just the integrand of the axial $\gamma W$-box diagram, which is an infrared-finite quantity; so it is more efficient to focus explicitly on the lattice calculation of this integrand.

\begin{figure}
	\begin{centering}
		\includegraphics[scale=0.6]{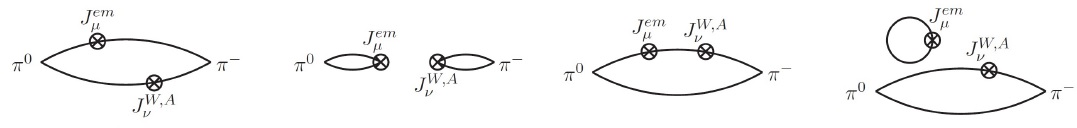}\hfill
		\par\end{centering}
	\caption{\label{fig:contraction}Contraction diagrams in the pion axial $\gamma W$-box. }
\end{figure}

The first realistic lattice calculation of such kind was performed on a simpler $\pi^+$--$\pi^0$ axial $\gamma W$-box diagram as a prototype~\cite{Feng:2020zdc,Seng:2021qdx}. Again, one first writes the box diagram correction to the $\pi_{e3}$ decay amplitude as:
\begin{equation}
\Box_\pi=\frac{3\alpha}{2\pi}\int \frac{dQ^2}{Q^2}\frac{M_W^2}{M_W^2+Q^2}\mathbb{M}_\pi(Q^2)~,
\end{equation}
where $\mathbb{M}_\pi(Q^2)$ denotes the first Nachtmann moment of the structure function $F_3^{(0)}$ of pion. Since pions are spinless, the axial box diagram does not receive a Born contribution, which makes the analysis much simpler.
At low $Q^2$, the function $\mathbb{M}_\pi(Q^2)$ was calculated on lattice by computing the four-point contraction diagrams depicted in Fig.\ref{fig:contraction}; meanwhile, at larger $Q^2$ such non-perturbative calculations are not applicable due to the increasing lattice artifacts, but pQCD results are available to $\mathcal{O}(\alpha_s^4)$~\cite{Baikov:2010je,Baikov:2010iw}. Furthermore, one observes a smooth connection between the pQCD curve and the lattice curve around $Q^2=2$~GeV$^2$, which provides further justifications of our previous assertion about the onset of the perturbative regime at $Q^2>2$~GeV$^2$.
In summary, by combining the perturbative and non-perturbative calculations, one obtained $\Box_\pi=2.830(11)_\mathrm{stat}(26)_\mathrm{syst}\times 10^{-3}$, which translates into $\delta_\pi=0.0332(3)$. The new result brings a three-fold improvement in precision comparing to the previous state-of-the-art calculation based on ChPT~\cite{Cirigliano:2002ng,Cirigliano:2003yr}.

Motivated by the successful first trial, a natural follow-up is to calculate the neutron axial $\gamma W$-box diagram directly on lattice. It could consist of the generalization of the aforementioned four-point function method, with possible extra complications:
\begin{itemize}
	\item The quark contraction becomes more complicated in the nucleon sector,
	\item The data is much noisier for nucleon due to the exponentially suppressed signal-to-noise ratio at large Euclidean time, thus much more computational resources are needed,
	\item The full control of the systematic effects such as the excited-state contamination becomes more challenging in the nucleon sector.
\end{itemize}
Notice, however, that the four-point function method is not the only way to proceed; a possible alternative is to make use of the Feynman-Hellmann theorem on lattice~\cite{Seng:2019plg}.

\subsection{Nuclear structure effects}

Next we proceed to the NS-dependent corrections in superallowed nuclear beta decays. Recall the master formula in Eq.\eqref{eq:Master0+}: The left hand side is nucleus-independent, and so must be the right hand side. But the experimentally-measured $ft$-values are nucleus-dependent, so one needs to carefully remove all the ``nucleus-dependent dressings'' in $ft$ to obtain a nucleus-independent $\mathcal{F}t$ value, which is usually expressed as~\cite{Hardy:2020qwl}:
\begin{equation}
\mathcal{F}t=ft(1+\delta_\mathrm{R}')(1+\delta_\mathrm{NS}-\delta_\mathrm{C})~.
\end{equation} 
There are three types of nuclear corrections:
\begin{enumerate}
	\item $\delta_\mathrm{R}'$: The extreme-infrared contribution of the RC (i.e. ``outer correction''), which is calculable to high precision~\cite{Sirlin:1967zza,Towner:2007np,Sirlin:1987sy,Sirlin:1986cc};
	\item $\delta_\mathrm{NS}$: The NS correction to the single-nucleon axial $\gamma W$-box diagram;
	\item $\delta_\mathrm{C}$: The strong ISB correction to the nuclear wavefunction.
\end{enumerate}

\begin{figure}
	\begin{centering}
		\includegraphics[scale=0.6]{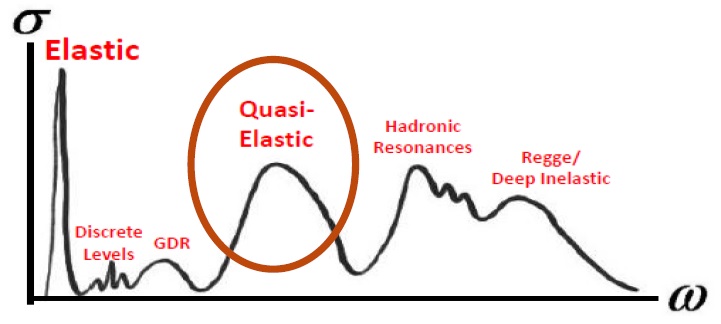}\hfill
		\par\end{centering}
	\caption{\label{fig:absorption}Idealized absorption spectrum of a nucleus. Circled is the quasi-elastic peak. }
\end{figure}

Nuclear models have been used to compute both $\delta_\mathrm{NS}$ and $\delta_\mathrm{C}$. Currently, the standard theory inputs for these quantities were taken from the calculations by Hardy and Towner based on shell model + Woods-Saxon (WS) potential; with that they reported $\overline{\mathcal{F}t}=3072.27(72)$s in their 2015 review~\cite{Hardy:2014qxa}. However, recent theory developments cast doubts on the validity of these calculations:
\begin{itemize}
	\item First, the correction $\delta_\mathrm{NS}$ entails the nuclear modifications to the single-nucleon $\gamma W$-box diagrams, in particular the Born contribution. In the Hardy-Towner treatment, the latter was evaluated by adding the so-called ``quenching factors'' to account for the observed reduction of the free-nucleon magnetic moment and the axial charge in the nuclear medium~\cite{Towner:1994mw,Towner:2002rg}. Such treatment was shown to be incomplete through the introduction of the dispersive representation to nuclear beta decays. By writing the nuclear $\gamma W$-box diagram in terms of single-current nuclear matrix elements, it is immediately apparent that the Hardy-Towner treatment has only accounted for the smaller contribution from the discrete energy levels at the lower end of the nuclear absorption spectrum, but has missed the large contribution from the broad quasi-elastic peak (see Fig.\ref{fig:absorption}), which affects the nuclear box diagram at both $E_e=0$ and $E_e\neq 0$, with $E_e$ the electron energy. Simple estimations based on a Fermi gas model~\cite{Seng:2018qru,Gorchtein:2018fxl} suggest that the $E_e=0$ and $E_e\neq 0$ corrections tend to partially cancel each other, leaving an enlarged nuclear uncertainty in $\delta_\mathrm{NS}$. With this update, Hardy and Towner quoted $\overline{\mathcal{F}t}=3072.24(1.85)$~s in their 2020 review~\cite{Hardy:2020qwl}, where the central value almost remains constant but the uncertainty is significantly larger than the 2015 result. 
	\item The ISB correction $\delta_\mathrm{C}$ plays a central role in obtaining a universal $\mathcal{F}t$-value for different superallowed transitions, which is required by CVC hypothesis. However, it turns out that among all existing nuclear theory calculations of $\delta_\mathrm{C}$, including shell-model with WS potential~\cite{Towner:2002rg,Towner:2007np,Hardy:2008gy}, Hartree-Fock wavefunctions~\cite{Ormand:1989hm,Ormand:1995df}, density functional theory~\cite{Satula:2011br}, random-phase approximation~\cite{Liang:2009pf}, isovector monopole resonance sum rule~\cite{Auerbach:2008ut} and the Damgaard model~\cite{Damgaard:1969yyx},
	only the WS calculation by Hardy and Towner
	was able to achieve such an alignment and was taken as the standard input for $\delta_\mathrm{C}$.
	Whether this indicates the success of this particular simplified nuclear picture, or rather the existence of some unexpected new physics in Nature (e.g. second-class currents~\cite{Weinberg:1958ut}), is not yet clear. Furthermore, questions have been raised on the internal consistency of the Hardy-Towner calculation~\cite{Miller:2008my,Miller:2009cg}.
\end{itemize}
In short, $\delta_\mathrm{NS}$ contributes the largest theory uncertainty in $V_{ud}$ at face value which must be reduced, whereas the apparent model-dependence in the current result of $\delta_\mathrm{C}$ needs to be understood.
Possible pathways for future improvements include  calculating the NS corrections with ab-initio methods, or constraining them (in particular the ISB corrections) model-independently using experimental measurements such as the neutron skin and charge radii~\cite{Koshchii:2020qkr}.

\section{$|V_{us}|$}

Following the original seminar, we discuss in the second half of this paper the extraction of $|V_{us}|$ from kaon decays. As we alerted before, there is currently a $\sim 2.8\sigma$ discrepancy between the value of $|V_{us}|$ extracted from the $K_{\mu 2}$ (i.e. $K\rightarrow\mu\nu(\gamma)$) and  $K_{\ell 3}$ (i.e. $K\rightarrow\pi\ell\nu(\gamma)$) decays of kaon:
\begin{equation}
|V_{us}|_{K_{\mu 2}}=0.2252(5)~,\:\:|V_{us}|_{K_{\ell 3}}=0.2231(6)~.
\label{eq:twoVus}
\end{equation}
Not only that the different choices of $|V_{us}|$ greatly affect the conclusions to the top-row CKM unitarity deficit, this discrepancy itself could be an indication of the existence of BSM physics. 
However, before making such conclusion, it is necessary to further scrutinize all the SM theory inputs in these decay processes to ensure that it is not due to some unexpected, large SM systematics. 

\subsection{From leptonic kaon decay}

From leptonic decays of kaon and pion one measures the following ratio~\cite{Marciano:2004uf,Cirigliano:2011tm}:
\begin{equation}
R_A\equiv\frac{\Gamma_{K_{\mu 2}}}{\Gamma_{\pi_{\mu 2}}}=\frac{|V_{us}|^2}{|V_{ud}|^2}\frac{f_{K^+}^2}{f_{\pi^+}^2}\frac{M_{K^+}(1-m_\mu^2/M_{K^+}^2)^2}{M_{\pi^+}(1-m_\mu^2/M_{\pi^+}^2)^2}(1+\delta_\mathrm{EM})~.
\end{equation}
Here $f_{K^+}$ and $f_{\pi^+}$ are the physical $K^+$ and $\pi^+$ decay constants (including the ISB corrections), whereas $\delta_\mathrm{EM}$ denotes the difference between the electromagnetic RC to $K_{\mu 2}$ and $\pi_{\mu 2}$. This gives us the ratio $|V_{us}/V_{ud}|$, where $|V_{ud}|$ is later taken from superallowed beta decays. Constructing the ratio $R_A$ results in the cancellation of some common theory uncertainties in $K_{\mu 2}$ and $\pi_{\mu 2}$ associated to (1) the universal short-distance RC, and (2) a part of the long-distance RC; the latter corresponds to the low energy constants (LECs) at $\mathcal{O}(e^2p^2)$ in the ChPT language. This allows us to determine $\delta_\mathrm{EM}$ theoretically to a higher precision than the individual RCs to $K_{\mu 2}$ and $\pi_{\mu 2}$ separately.

In obtaining the quoted value of $|V_{us}|_{K_{\mu 2}}$ in Eq.\eqref{eq:twoVus}, we made use of the ratio of the decay constants from the 2021 FLAG review at $N_f=2+1+1$:
$f_{K^+}/f_{\pi^+}=1.1932(21)$~\cite{Aoki:2021kgd} (averaging over Refs.\cite{Bazavov:2017lyh,Dowdall:2013rya,Carrasco:2014poa,Miller:2020xhy}), as well as the ChPT result of the long-distance electromagnetic RC: $\delta_\mathrm{EM}=-0.0069(17)$~\cite{Cirigliano:2011tm}. Recently there are direct lattice QCD calculations of the electromagnetic and ISB corrections to the ratio of the $K_{\mu 2}$ and $\pi_{\mu 2}$ decay rates~\cite{Giusti:2017dwk,DiCarlo:2019thl}; the outcomes are consistent with the ChPT result with comparable levels of precision, which provide strong support to the reliability of the theory inputs in these processes.

\subsection{From semileptonic kaon decays} 

$|V_{us}|$ can also be extracted from semileptonic kaon decays through the following master formula:
\begin{equation}
\Gamma_{K_{\ell 3}}=\frac{G_F^2|V_{us}|^2M_K^5C_K^2}{192\pi^3}S_\mathrm{EW}|f_+^{K^0\pi^-}(0)|^2I_{K\ell}^{(0)}\left(1+\delta_\mathrm{EM}^{K\ell}+\delta_\mathrm{SU(2)}^{K\pi}\right)~.
\end{equation}
There are six independent modes in $K_{\ell 3}$ decays ($K_{e3}^\pm$, $K_{e3}^S$, $K_{e3}^L$, $K_{\mu 3}^\pm$, $K_{\mu 3}^S$, $K_{\mu 3}^L$), and the required experimental inputs are the lifetimes and the branching ratios. Measurements of the branching ratios have been done in all six modes~\cite{KLOE:2005vdt,KTeV:2004hpx,Batley:2007zzb,KLOE:2006vvm,KLOE:2002lao,KLOE:2007jte,Chiang:1972rp}, which allow us to average over different modes in order to reduce the experimental uncertainties in $|V_{us}|$.

There are several theory inputs needed in the master formula above, and the only trivial one is the isospin factor $C_K$ which equals $1$ for $K^0$ and $1/\sqrt{2}$ for $K^+$. In what follows we briefly discuss the meanings and the current status of the remaining, non-trivial theory inputs.

\subsubsection{$S_\mathrm{EW}$:} 
It denotes the short-distance electroweak RC which is universal to all semileptonic/leptonic beta decays~\cite{Marciano:1993sh}. It contains the large electroweak logarithm, the leading pQCD corrections and the resummed large QED logs, and is often presented schematically as~\cite{Cirigliano:2008wn,Cirigliano:2011ny}:
	\begin{equation}
	S_\mathrm{EW}=1+\frac{2\alpha}{\pi}\left(1-\frac{\alpha_s}{4\pi}\right)\ln\frac{M_Z}{M_\rho}+\mathcal{O}\left(\frac{\alpha\alpha_s}{\pi^2}\right)~,
	\end{equation}
where the $\rho$-mass appears as a conventionally-chosen low-energy scale. The numerical value $S_\mathrm{EW}=1.0232(3)$ is always adopted for all practical purposes~\cite{Cirigliano:2001mk,Cirigliano:2004pv,Antonelli:2010yf}, and the associated theory uncertainty is the smallest among all theory inputs.

\subsubsection{$f_+^{K^0\pi^-}(0)$:} 

This is the $K^0\rightarrow\pi^-$ charged weak form factor at the (unphysical) zero-momentum-transfer point, defined through:
	\begin{equation}
	\left\langle \pi^-(p_\pi)\right|J_W^\mu\left|K^0(p_K)\right\rangle=f_+^{K^0\pi^-}(t)(p_K+p_\pi)^\mu+f_-^{K^0\pi^-}(t)(p_K-p_\pi)^\mu~,\label{eq:formfactors}
	\end{equation}
with $t=(p_K-p_\pi)^2$. 
The quoted result of $|V_{us}|_{K_{\ell 3}}$ in Eq.\eqref{eq:twoVus} has made use of the latest FLAG average for the $N_f=2+1+1$ lattice results: $|f_+^{K^0\pi^-}(0)|=0.9698(17)$~\cite{Aoki:2021kgd} (averaging over Refs.\cite{Carrasco:2016kpy,Bazavov:2018kjg}). However, a new lattice calculation by the PACS Collaboration at $N_f=2+1$ with two lattice spacings return a smaller value, $|f_+^{K^0\pi^-}(0)|=0.9605(39)(27)$~\cite{Kakazu:2019ltq,Yamazaki:2021zxz,PACStalk}. Given that the $K_{\mu 2}$--$K_{\ell 3}$ discrepancy would largely diminish should the PACS result be used, the reason of such a disagreement must be properly understood.

\subsubsection{$I_{K\ell}^{(0)}$:} 
This is the $K_{\ell 3}$ phase space factor which probes the $t$-dependence of the $K\pi$ form factors $f_\pm^{K\pi}(t)$. The latter is obtained by fitting to the $K_{\ell 3}$ Dalitz plot assuming a specific parameterization; examples are the Taylor expansion~\cite{Antonelli:2010yf},
the $z$-parameterization~\cite{Hill:2006bq}, the pole parameterization~\cite{Lichard:1997ya} and
the dispersive parameterization~\cite{Bernard:2006gy,Bernard:2009zm,Abouzaid:2009ry}. One of the most widely-adopted choices is the dispersive parameterization, with the theory uncertainty controlled at a (0.10--0.15)\% level~\cite{MoulsonCKM21}.

\subsubsection{$\delta_\mathrm{EM}^{K\ell}$:}
 
This represents the long-distance RC which is not already contained in $S_\mathrm{EW}$. The study of this RC is somewhat more challenging than that of $\pi_{e3}$ and free neutron decay, because the $K\pi$ mass splitting is not small.
In the past two decades, the standard inputs for this correction were taken from ChPT calculations~\cite{Cirigliano:2001mk,Cirigliano:2003yr,Cirigliano:2004pv,Cirigliano:2008wn}. Two major sources of theory uncertainty within this framework are: (1) The neglected higher-order terms at $\mathcal{O}(e^2p^4)$, which effects were estimated using simple chiral power counting, and (2) The poorly-known LECs at $\mathcal{O}(e^2p^2)$, which 
were estimated within resonance models~\cite{Ananthanarayan:2004qk,DescotesGenon:2005pw}. Both of them led to theory uncertainties at the order $10^{-3}$, which signifies the ``natural limitation'' of the ChPT framework. There are also plans to calculate the fully inclusive RC in $K_{\ell 3}$ using the same lattice QCD method previously applied to $K_{\mu 2}$, but it is expected to take up to $\sim$10 years to reach a $10^{-3}$ precision~\cite{BoyleSnowmass}.

A recent reevaluation of the electromagnetic RC in $K_{e3}$ using a new theory framework~\cite{Seng:2019lxf,Seng:2020jtz} based on the hybridization of the classical Sirlin's representation of RC (i.e. that we described at the beginning of Sec.\ref{sec:singlenucleon} ), ChPT and lattice QCD has successfully overcome the aforementioned natural limitation. The new framework allows one to resum the most important higher-order ChPT corrections, and to utilize the most recent lattice QCD calculation of the $K\pi$ axial $\gamma W$-box diagram in the SU(3) limit to effectively reduce the uncertainties from the LECs~\cite{Ma:2021azh}. The new results agree with the ChPT prediction of the $K_{e3}$ RC, but improves its precision from $10^{-3}$ to $10^{-4}$~\cite{Seng:2021boy,Seng:2021wcf}. This indicates that the long-distance RC is very unlikely to be the underlying reason for the $K_{\ell 3}$--$K_{\mu 2}$ discrepancy.

\subsubsection{$\delta_\mathrm{SU(2)}^{K\pi}$:}
This represents the ISB correction to the $K\pi$ form factor at zero momentum transfer, and is rigorously defined as\footnote{Here we adopt the normalization convention $|f_+^{K\pi}(0)|=C_K$ in the SU(3) limit, in consistency with Eq.\eqref{eq:formfactors}.}:
\begin{equation}
\delta_\mathrm{SU(2)}^{K\pi}=\left(\frac{C_{K^0}}{C_K}\frac{f_+^{K\pi}(0)}{f_+^{K^0\pi^-}(0)}\right)^2-1
\end{equation}
which only exists in the $K^+$ channel by construction. Upon neglecting small electromagnetic contributions, this correction is fixed by the quark mass parameters $m_s/\hat{m}$ and $\mathbb{Q}^2\equiv (m_s^2-\hat{m}^2)/(m_d^2-m_u^2)$, where $\hat{m}=(m_u+m_d)/2$~\cite{Antonelli:2010yf}. These parameters can either be obtained from lattice QCD~\cite{RBC:2014ntl,Durr:2010vn,Durr:2010aw,MILC:2009ltw,Fodor:2016bgu} or from the phenomenological analysis of $\eta\rightarrow 3\pi$~\cite{Colangelo:2018jxw}. The former gives $\delta_\mathrm{SU(2)}^{K^+\pi^0}=0.0457(20)$~\cite{Seng:2021nar}, while the latter gives a somewhat larger value of $\delta_\mathrm{SU(2)}^{K^+\pi^0}=0.0522(34)$~\cite{MoulsonCKM21}. The discrepancy between these two determinations needs be better understood.

\section{A short summary}

The precision frontier plays an important role in the search of BSM physics, and one of the latest observed anomalies at this frontier is the so-called Cabibbo angle anomaly, which includes an apparent $(2-3)\sigma$ deficit from the top-row CKM matrix unitarity through the precise determinations of $|V_{ud}|$ and $|V_{us}|$ from various beta decay experiments. In order to confirm this finding, one needs to further reduce all the SM theory uncertainties in these matrix elements. 

The past few years have seen a number of promising progress in the understanding of the single-nucleon RC in free neutron and superallowed nuclear beta decays through the introduction of the DR treatment, which efficiently constrains the non-perturbative QCD at the hadron scale using experimental data. Furthermore, a near-future lattice QCD calculation --- with a successful prototype in the pion sector --- may fully pin down this correction. 
On the other hand, several inconsistencies in the existing NS corrections to superallowed beta decays have been pointed out and require further scrutiny. Ab-initio calculations and independent constraints from experiments are highly desirable.

Finally, a $\sim 2.8\sigma$ disagreement between the values of $|V_{us}|$ measured from different channels of kaon decays is also observed, which calls for an immediate resolution. In contrast to those in the $K_{\mu 2}$ decay which are relative under control, there are several SM theory inputs in $K_{\ell 3}$ that worth a closer look. A recent reevaluation of the long-distance RC in $K_{e3}$ confirms the previous calculation with higher precision, but on the other hand discrepancies exist between the recent $N_f=2+1$ determination of $f_+^{K^0\pi^-}(0)$ and the $N_f=2+1+1$ averages, and between the lattice and phenomenological determinations of the ISB corrections. Careful cross-checks are needed in order to remove all possible underlying SM systematic effects from these inputs. 

\section*{Acknowledgments}

The author expresses his gratitude to Bastian M\"{a}rkisch for important clarifications on the current experimental situation of the neutron decay, and also thanks Jens Erler, Xu Feng, Daniel Galviz, Mikhail Gorchtein, Charles J. Horowitz, Lu-Chang Jin, Oleksandr Koshchii, Peng-Xiang Ma, William J. Marciano, Ulf-G. Mei{\ss}ner, Hiren H. Patel, Jorge Piekarewicz, Michael J. Ramsey-Musolf, Xavier Roca-Maza and Hubert Spiesberger for collaborations in related topics. 
This work is supported in
part by the Deutsche Forschungsgemeinschaft (DFG, German Research
Foundation) and the NSFC through the funds provided to the Sino-German Collaborative Research Center TRR110 “Symmetries and the Emergence of Structure in QCD” (DFG Project-ID 196253076 - TRR 110, NSFC Grant No. 12070131001).

\bibliographystyle{JHEP-2}
\bibliography{Ke3_ref}

\providecommand{\href}[2]{#2}\begingroup\raggedright\begin{thebibliography}{100}

\bibitem{RikenBNL}
url: \url{https://indico.bnl.gov/event/9317/?print=1}.

\bibitem{Aghanim:2018eyx}
{\bf Planck} Collaboration, N.~Aghanim {\em et.~al.}, {\it {Planck 2018
  results. VI. Cosmological parameters}},  {\em Astron. Astrophys.} {\bf 641}
  (2020) A6 [\href{http://arXiv.org/abs/1807.06209}{{\tt 1807.06209}}].

\bibitem{Riess:1998cb}
{\bf Supernova Search Team} Collaboration, A.~G. Riess {\em et.~al.}, {\it
  {Observational evidence from supernovae for an accelerating universe and a
  cosmological constant}},  {\em Astron. J.} {\bf 116} (1998) 1009--1038
  [\href{http://arXiv.org/abs/astro-ph/9805201}{{\tt astro-ph/9805201}}].

\bibitem{Perlmutter:1998np}
{\bf Supernova Cosmology Project} Collaboration, S.~Perlmutter {\em et.~al.},
  {\it {Measurements of $\Omega$ and $\Lambda$ from 42 high redshift
  supernovae}},  {\em Astrophys. J.} {\bf 517} (1999) 565--586
  [\href{http://arXiv.org/abs/astro-ph/9812133}{{\tt astro-ph/9812133}}].

\bibitem{Simon:2019nxf}
J.~D. Simon, {\it {The Faintest Dwarf Galaxies}},  {\em Ann. Rev. Astron.
  Astrophys.} {\bf 57} (2019), no.~1 375--415
  [\href{http://arXiv.org/abs/1901.05465}{{\tt 1901.05465}}].

\bibitem{Salucci:2018hqu}
P.~Salucci, {\it {The distribution of dark matter in galaxies}},  {\em Astron.
  Astrophys. Rev.} {\bf 27} (2019), no.~1 2
  [\href{http://arXiv.org/abs/1811.08843}{{\tt 1811.08843}}].

\bibitem{Allen:2011zs}
S.~W. Allen, A.~E. Evrard and A.~B. Mantz, {\it {Cosmological Parameters from
  Observations of Galaxy Clusters}},  {\em Ann. Rev. Astron. Astrophys.} {\bf
  49} (2011) 409--470 [\href{http://arXiv.org/abs/1103.4829}{{\tt 1103.4829}}].

\bibitem{Sakharov:1967dj}
A.~D. Sakharov, {\it {Violation of CP Invariance, C asymmetry, and baryon
  asymmetry of the universe}},  {\em Sov. Phys. Usp.} {\bf 34} (1991), no.~5
  392--393.

\bibitem{Mossa:2020gjc}
V.~Mossa {\em et.~al.}, {\it {The baryon density of the Universe from an
  improved rate of deuterium burning}},  {\em Nature} {\bf 587} (2020),
  no.~7833 210--213.

\bibitem{Fermigm2}
{\bf Muon g-2} Collaboration, B.~Abi {\em et.~al.}, {\it {Measurement of the
  Positive Muon Anomalous Magnetic Moment to 0.46 ppm}},  {\em Phys. Rev.
  Lett.} {\bf 126} (2021) 141801.

\bibitem{Aoyama:2020ynm}
T.~Aoyama {\em et.~al.}, {\it {The anomalous magnetic moment of the muon in the
  Standard Model}},  {\em Phys. Rept.} {\bf 887} (2020) 1--166
  [\href{http://arXiv.org/abs/2006.04822}{{\tt 2006.04822}}].

\bibitem{Miller:2007kk}
J.~P. Miller, E.~de~Rafael and B.~L. Roberts, {\it {Muon (g-2): Experiment and
  theory}},  {\em Rept. Prog. Phys.} {\bf 70} (2007) 795
  [\href{http://arXiv.org/abs/hep-ph/0703049}{{\tt hep-ph/0703049}}].

\bibitem{Miller:2012opa}
J.~P. Miller, E.~de~Rafael, B.~L. Roberts and D.~St\"ockinger, {\it {Muon
  (g-2): Experiment and Theory}},  {\em Ann. Rev. Nucl. Part. Sci.} {\bf 62}
  (2012) 237--264.

\bibitem{Jegerlehner:2009ry}
F.~Jegerlehner and A.~Nyffeler, {\it {The Muon g-2}},  {\em Phys. Rept.} {\bf
  477} (2009) 1--110 [\href{http://arXiv.org/abs/0902.3360}{{\tt 0902.3360}}].

\bibitem{Aaij:2019wad}
{\bf LHCb} Collaboration, R.~Aaij {\em et.~al.}, {\it {Search for
  lepton-universality violation in $B^+\to K^+\ell^+\ell^-$ decays}},  {\em
  Phys. Rev. Lett.} {\bf 122} (2019), no.~19 191801
  [\href{http://arXiv.org/abs/1903.09252}{{\tt 1903.09252}}].

\bibitem{Aaij:2014ora}
{\bf LHCb} Collaboration, R.~Aaij {\em et.~al.}, {\it {Test of lepton
  universality using $B^{+}\rightarrow K^{+}\ell^{+}\ell^{-}$ decays}},  {\em
  Phys. Rev. Lett.} {\bf 113} (2014) 151601
  [\href{http://arXiv.org/abs/1406.6482}{{\tt 1406.6482}}].

\bibitem{Aaij:2015yra}
{\bf LHCb} Collaboration, R.~Aaij {\em et.~al.}, {\it {Measurement of the ratio
  of branching fractions $\mathcal{B}(\bar{B}^0 \to
  D^{*+}\tau^{-}\bar{\nu}_{\tau})/\mathcal{B}(\bar{B}^0 \to
  D^{*+}\mu^{-}\bar{\nu}_{\mu})$}},  {\em Phys. Rev. Lett.} {\bf 115} (2015),
  no.~11 111803 [\href{http://arXiv.org/abs/1506.08614}{{\tt 1506.08614}}].
  [Erratum: Phys.Rev.Lett. 115, 159901 (2015)].

\bibitem{Aaij:2015oid}
{\bf LHCb} Collaboration, R.~Aaij {\em et.~al.}, {\it {Angular analysis of the
  $B^{0} \to K^{*0} \mu^{+} \mu^{-}$ decay using 3 fb$^{-1}$ of integrated
  luminosity}},  {\em JHEP} {\bf 02} (2016) 104
  [\href{http://arXiv.org/abs/1512.04442}{{\tt 1512.04442}}].

\bibitem{Hardy:2020qwl}
J.~C. Hardy and I.~S. Towner, {\it {Superallowed $0^+ \to 0^+$ nuclear $\beta$
  decays: 2020 critical survey, with implications for V$_{ud}$ and CKM
  unitarity}},  {\em Phys. Rev. C} {\bf 102} (2020), no.~4 045501.

\bibitem{Zyla:2020zbs}
{\bf Particle Data Group} Collaboration, P.~Zyla {\em et.~al.}, {\it {Review of
  Particle Physics}},  {\em PTEP} {\bf 2020} (2020), no.~8 083C01.

\bibitem{Hayen:2020cxh}
L.~Hayen, {\it {Standard model $\mathcal{O}(\alpha)$ renormalization of $g_A$
  and its impact on new physics searches}},  {\em Phys. Rev. D} {\bf 103}
  (2021), no.~11 113001 [\href{http://arXiv.org/abs/2010.07262}{{\tt
  2010.07262}}].

\bibitem{Feng:2020zdc}
X.~Feng, M.~Gorchtein, L.-C. Jin, P.-X. Ma and C.-Y. Seng, {\it
  {First-principles calculation of electroweak box diagrams from lattice QCD}},
   {\em Phys. Rev. Lett.} {\bf 124} (2020), no.~19 192002
  [\href{http://arXiv.org/abs/2003.09798}{{\tt 2003.09798}}].

\bibitem{Seng:2021nar}
C.-Y. Seng, D.~Galviz, W.~J. Marciano and U.-G. Mei\ss{}ner, {\it {An update on
  $|V_{us}|$ and $|V_{us}/V_{ud}|$ from semileptonic kaon and pion decays}},
  \href{http://arXiv.org/abs/2107.14708}{{\tt 2107.14708}}.

\bibitem{HFLAV:2019otj}
{\bf HFLAV} Collaboration, Y.~S. Amhis {\em et.~al.}, {\it {Averages of
  b-hadron, c-hadron, and $\tau $-lepton properties as of 2018}},  {\em Eur.
  Phys. J. C} {\bf 81} (2021), no.~3 226
  [\href{http://arXiv.org/abs/1909.12524}{{\tt 1909.12524}}].

\bibitem{Cabibbo:2003ea}
N.~Cabibbo, E.~C. Swallow and R.~Winston, {\it {Semileptonic hyperon decays and
  CKM unitarity}},  {\em Phys. Rev. Lett.} {\bf 92} (2004) 251803
  [\href{http://arXiv.org/abs/hep-ph/0307214}{{\tt hep-ph/0307214}}].

\bibitem{Pauli:1949zm}
W.~Pauli and F.~Villars, {\it {On the Invariant regularization in relativistic
  quantum theory}},  {\em Rev. Mod. Phys.} {\bf 21} (1949) 434--444.
%%CITATION = RMPHA,21,434;%%

\bibitem{Wu:1957my}
C.~S. Wu, E.~Ambler, R.~W. Hayward, D.~D. Hoppes and R.~P. Hudson, {\it
  {Experimental Test of Parity Conservation in $\beta$ Decay}},  {\em Phys.
  Rev.} {\bf 105} (1957) 1413--1414.

\bibitem{Lee:1956qn}
T.~D. Lee and C.-N. Yang, {\it {Question of Parity Conservation in Weak
  Interactions}},  {\em Phys. Rev.} {\bf 104} (1956) 254--258.

\bibitem{Feynman:1958ty}
R.~P. Feynman and M.~Gell-Mann, {\it {Theory of Fermi interaction}},  {\em
  Phys. Rev.} {\bf 109} (1958) 193--198.

\bibitem{Sudarshan:1958vf}
E.~C.~G. Sudarshan and R.~e. Marshak, {\it {Chirality invariance and the
  universal Fermi interaction}},  {\em Phys. Rev.} {\bf 109} (1958) 1860--1860.

\bibitem{Cabibbo:1963yz}
N.~Cabibbo, {\it {Unitary Symmetry and Leptonic Decays}},  {\em Phys. Rev.
  Lett.} {\bf 10} (1963) 531--533.
%%CITATION = PRLTA,10,531;%%

\bibitem{Kobayashi:1973fv}
M.~Kobayashi and T.~Maskawa, {\it {CP Violation in the Renormalizable Theory of
  Weak Interaction}},  {\em Prog. Theor. Phys.} {\bf 49} (1973) 652--657.
%%CITATION = PTPKA,49,652;%%

\bibitem{Crivellin:2020lzu}
A.~Crivellin and M.~Hoferichter, {\it {\ensuremath{\beta} Decays as Sensitive
  Probes of Lepton Flavor Universality}},  {\em Phys. Rev. Lett.} {\bf 125}
  (2020), no.~11 111801 [\href{http://arXiv.org/abs/2002.07184}{{\tt
  2002.07184}}].

\bibitem{Crivellin:2021njn}
A.~Crivellin, M.~Hoferichter and C.~A. Manzari, {\it {Fermi Constant from Muon
  Decay Versus Electroweak Fits and Cabibbo-Kobayashi-Maskawa Unitarity}},
  {\em Phys. Rev. Lett.} {\bf 127} (2021), no.~7 071801
  [\href{http://arXiv.org/abs/2102.02825}{{\tt 2102.02825}}].

\bibitem{UMass18}
\textit{Beta decay as a probe of new physics}, UMass Amherst, November 2018.
  [\href{https://www.physics.umass.edu/acfi/seminars-and-workshops/beta-decay-as-a-probe-of-new-physics}{{\tt
  Website}}].

\bibitem{TAMU19}
\textit{Top row CKM unitarity workshop}, Texas A\&M University, January 2019.
  [\href{https://cyclotron.tamu.edu/ckmuw2019}{{\tt Website}}].

\bibitem{ECT19}
\textit{Precise beta decay calculations for searches for new physics}, ECT*,
  April 2019. [\href{https://indico.ectstar.eu/event/42}{{\tt Website}}].

\bibitem{UMass19}
\textit{Current and future status of the first-row CKM unitarity}, UMass
  Amherst, May 2019.
  [\href{https://www.physics.umass.edu/acfi/seminars-and-workshops/current-and-future-status-of-the-first-row-ckm-unitarity}{{\tt
  Website}}].

\bibitem{INT19}
\textit{Fundamental symmetries research with beta decay}, INT, University of
  Washington, November 2019.
  [\href{https://www.int.washington.edu/PROGRAMS/19-75W}{{\tt Website}}].

\bibitem{APS20}
\textit{Q04 invited session: BSM physics in nuclear and neutron beta decay},
  APS Virtual April Meeting, April 2020.
  [\href{http://meetings.aps.org/Meeting/APR20/Session/Q04}{{\tt Website}}].

\bibitem{MITP20}
\textit{Parity violation and related topics}, MITP, Johannes Gutenberg
  University of Mainz, July 2020.
  [\href{https://indico.mitp.uni-mainz.de/event/217}{{\tt Website}}].

\bibitem{Snowmass20}
\textit{RF2: weak decays of strange and light quarks}, Rare processes and
  precision frontier townhall meeting, October 2020.
  [\href{https://indico.fnal.gov/event/45713/sessions/16417/#20201002}{{\tt
  Website}}].

\bibitem{Moriond21}
Moriond 2021, Electroweak Interactions and Unified Theories.
  [\href{https://moriond.in2p3.fr/2021/EW/}{{\tt Website}}].

\bibitem{DPG21}
Deutschen Physikalischen Gesellschaft (German Physical Society, DPG) Meeting of
  the Matter and Cosmos Section (SMuK), 2021.
  [\href{https://smuk21.dpg-tagungen.de/}{{\tt Website}}].

\bibitem{DNP21}
Fall Meeting of the Division of Nuclear Physics (DNP) of the American Physical
  Society (APS), 2021. [\href{http://web.mit.edu/dnp2021/}{{\tt Website}}].

\bibitem{CKM21}
11th International Workshop on the CKM Unitarity Triangle (CKM 2021).
  [\href{https://indico.cern.ch/event/891123/}{{\tt Website}}].

\bibitem{BaccaSnowmass}
S.~Bacca {\em et.~al.}
\newblock \textit{The quest for explaining the top-row CKM unitarity deficit}.
  [\href{https://www.snowmass21.org/docs/files/summaries/RF/SNOWMASS21-RF2_RF3-NF0_NF0-TF0_TF0_Gorshteyn-055.pdf}{{\tt
  Link}}].

\bibitem{BaesslerSnowmass}
S.~Baessler {\em et.~al.}
\newblock \textit{Neutron beta decay in the test of the Unitarity of the CKM
  matrix}.
  [\href{https://www.snowmass21.org/docs/files/summaries/RF/SNOWMASS21-RF0_RF3-102.pdf}{{\tt
  Link}}].

\bibitem{ArevaloSnowmass}
A.~Aguilar-Arevalo {\em et.~al.}
\newblock \textit{Testing Lepton Flavor Universality and CKM Unitarity with
  Rare Pion Decay}.
  [\href{https://www.snowmass21.org/docs/files/summaries/RF/SNOWMASS21-RF2_RF3-048.pdf}{{\tt
  Link}}].

\bibitem{BazavovSnowmass}
A.~Bazavov {\em et.~al.}
\newblock \textit{Precise Lattice QCD calculations of kaon and pion decay
  parameters and first-row CKM unitarity tests}.
  [\href{https://www.snowmass21.org/docs/files/summaries/RF/SNOWMASS21-RF2_RF0-EF5_EF0-TF5_TF0-CompF2_CompF0_El-Khadra-094.pdf}{{\tt
  Link}}].

\bibitem{BoyleSnowmass}
P.~Boyle {\em et.~al.}
\newblock \textit{High-precision determination of $V_{us}$ and $V_{ud}$ from
  lattice QCD}.
  [\href{https://www.snowmass21.org/docs/files/summaries/RF/SNOWMASS21-RF2_RF0-TF5_TF0-CompF2_CompF0-054.pdf}{{\tt
  Link}}].

\bibitem{BhattacharyaSnowmass}
T.~Bhattacharya {\em et.~al.}
\newblock \textit{Unitarity of CKM Matrix, $|V_{ud}|$, Radiative Corrections
  and Semi-leptonic Form Factors}.
  [\href{https://www.snowmass21.org/docs/files/summaries/EF/SNOWMASS21-EF4_EF5-RF2_RF3_Rajan_Gupta-249.pdf}{{\tt
  Link}}].

\bibitem{MuLan:2012sih}
{\bf MuLan} Collaboration, V.~Tishchenko {\em et.~al.}, {\it {Detailed Report
  of the MuLan Measurement of the Positive Muon Lifetime and Determination of
  the Fermi Constant}},  {\em Phys. Rev. D} {\bf 87} (2013), no.~5 052003
  [\href{http://arXiv.org/abs/1211.0960}{{\tt 1211.0960}}].

\bibitem{Behrends:1960nf}
R.~E. Behrends and A.~Sirlin, {\it {Effect of mass splittings on the conserved
  vector current}},  {\em Phys. Rev. Lett.} {\bf 4} (1960) 186--187.

\bibitem{Ademollo:1964sr}
M.~Ademollo and R.~Gatto, {\it {Nonrenormalization Theorem for the Strangeness
  Violating Vector Currents}},  {\em Phys. Rev. Lett.} {\bf 13} (1964)
  264--265.

\bibitem{Sirlin:1977sv}
A.~Sirlin, {\it {Current Algebra Formulation of Radiative Corrections in Gauge
  Theories and the Universality of the Weak Interactions}},  {\em Rev. Mod.
  Phys.} {\bf 50} (1978) 573. [Erratum: Rev. Mod. Phys.50,905(1978)].
%%CITATION = RMPHA,50,573;%%

\bibitem{Cirigliano:2002ng}
V.~Cirigliano, M.~Knecht, H.~Neufeld and H.~Pichl, {\it {The Pionic beta decay
  in chiral perturbation theory}},  {\em Eur. Phys. J. C} {\bf 27} (2003)
  255--262 [\href{http://arXiv.org/abs/hep-ph/0209226}{{\tt hep-ph/0209226}}].

\bibitem{Cirigliano:2003yr}
V.~Cirigliano, {\it {K(e3) and pi(e3) decays: Radiative corrections and CKM
  unitarity}},  in {\em {38th Rencontres de Moriond on Electroweak Interactions
  and Unified Theories}}, 5, 2003.
\newblock \href{http://arXiv.org/abs/hep-ph/0305154}{{\tt hep-ph/0305154}}.

\bibitem{Pocanic:2003pf}
D.~Pocanic {\em et.~al.}, {\it {Precise measurement of the pi+
  ---\ensuremath{>} pi0 e+ nu branching ratio}},  {\em Phys. Rev. Lett.} {\bf
  93} (2004) 181803 [\href{http://arXiv.org/abs/hep-ex/0312030}{{\tt
  hep-ex/0312030}}].

\bibitem{Czarnecki:2019iwz}
A.~Czarnecki, W.~J. Marciano and A.~Sirlin, {\it {Pion beta decay and
  Cabibbo-Kobayashi-Maskawa unitarity}},  {\em Phys. Rev. D} {\bf 101} (2020),
  no.~9 091301 [\href{http://arXiv.org/abs/1911.04685}{{\tt 1911.04685}}].

\bibitem{Pioneer}
D.~Hertzog, 2021.
\newblock A next-generation rare pion decay experiment to study LFUV and CKM
  unitarity, \textit{The 16th International Workshop on Tau Lepton Physics
  (TAU2021)},
  \url{https://indico.cern.ch/event/848732/contributions/4507273/attachments/2317723/3947483/Hertzog-TauLepton-2021.pdf}.

\bibitem{UCNt:2021pcg}
{\bf UCN\ensuremath{\tau}} Collaboration, F.~M. Gonzalez {\em et.~al.}, {\it
  {Improved Neutron Lifetime Measurement with UCN\ensuremath{\tau}}},  {\em
  Phys. Rev. Lett.} {\bf 127} (2021), no.~16 162501
  [\href{http://arXiv.org/abs/2106.10375}{{\tt 2106.10375}}].

\bibitem{Fornal:2018eol}
B.~Fornal and B.~Grinstein, {\it {Dark Matter Interpretation of the Neutron
  Decay Anomaly}},  {\em Phys. Rev. Lett.} {\bf 120} (2018), no.~19 191801
  [\href{http://arXiv.org/abs/1801.01124}{{\tt 1801.01124}}]. [Erratum:
  Phys.Rev.Lett. 124, 219901 (2020)].

\bibitem{Markisch:2018ndu}
B.~M\"arkisch {\em et.~al.}, {\it {Measurement of the Weak Axial-Vector
  Coupling Constant in the Decay of Free Neutrons Using a Pulsed Cold Neutron
  Beam}},  {\em Phys. Rev. Lett.} {\bf 122} (2019), no.~24 242501
  [\href{http://arXiv.org/abs/1812.04666}{{\tt 1812.04666}}].

\bibitem{Beck:2019xye}
M.~Beck {\em et.~al.}, {\it {Improved determination of the
  $\beta$-$\overline{\nu}_e$ angular correlation coefficient $a$ in free
  neutron decay with the $aSPECT$ spectrometer}},  {\em Phys. Rev. C} {\bf 101}
  (2020), no.~5 055506 [\href{http://arXiv.org/abs/1908.04785}{{\tt
  1908.04785}}].

\bibitem{Holstein:1974zf}
B.~R. Holstein, {\it {Recoil Effects in Allowed beta Decay: The Elementary
  Particle Approach}},  {\em Rev. Mod. Phys.} {\bf 46} (1974) 789. [Erratum:
  Rev.Mod.Phys. 48, 673--673 (1976)].

\bibitem{Wilkinson:1982hu}
D.~H. Wilkinson, {\it {ANALYSIS OF NEUTRON BETA DECAY}},  {\em Nucl. Phys. A}
  {\bf 377} (1982) 474--504.

\bibitem{Gudkov:2008pf}
V.~P. Gudkov, {\it {Asymmetry of recoil protons in neutron beta-decay}},  {\em
  Phys. Rev. C} {\bf 77} (2008) 045502
  [\href{http://arXiv.org/abs/0801.4896}{{\tt 0801.4896}}].

\bibitem{Ivanov:2012qe}
A.~N. Ivanov, M.~Pitschmann and N.~I. Troitskaya, {\it {Neutron $\beta^-$decay
  as a laboratory for testing the standard model}},  {\em Phys. Rev. D} {\bf
  88} (2013), no.~7 073002 [\href{http://arXiv.org/abs/1212.0332}{{\tt
  1212.0332}}].

\bibitem{Ivanov:2020ybx}
A.~N. Ivanov, R.~H\"ollwieser, N.~I. Troitskaya, M.~Wellenzohn and Y.~A.
  Berdnikov, {\it {Corrections of order $O(E^2_e/m^2_N)$, caused by weak
  magnetism and proton recoil, to the neutron lifetime and correlation
  coefficients of the neutron beta decay}},  {\em Results Phys.} {\bf 21}
  (2021) 103806 [\href{http://arXiv.org/abs/2010.14336}{{\tt 2010.14336}}].

\bibitem{Seng:2021syx}
C.-Y. Seng, {\it {Radiative corrections to semileptonic beta decays: Progress
  and challenges}},  \href{http://arXiv.org/abs/2108.03279}{{\tt 2108.03279}}.

\bibitem{Marciano:1982mm}
W.~J. Marciano and A.~Sirlin, {\it {RADIATIVE CORRECTIONS TO ATOMIC PARITY
  VIOLATION}},  {\em Phys. Rev. D} {\bf 27} (1983) 552.

\bibitem{Marciano:1983ss}
W.~J. Marciano and A.~Sirlin, {\it {On Some General Properties of the O(alpha)
  Corrections to Parity Violation in Atoms}},  {\em Phys. Rev. D} {\bf 29}
  (1984) 75. [Erratum: Phys.Rev.D 31, 213 (1985)].

\bibitem{Czarnecki:2004cw}
A.~Czarnecki, W.~J. Marciano and A.~Sirlin, {\it {Precision measurements and
  CKM unitarity}},  {\em Phys. Rev. D} {\bf 70} (2004) 093006
  [\href{http://arXiv.org/abs/hep-ph/0406324}{{\tt hep-ph/0406324}}].

\bibitem{Marciano:2005ec}
W.~J. Marciano and A.~Sirlin, {\it {Improved calculation of electroweak
  radiative corrections and the value of V(ud)}},  {\em Phys. Rev. Lett.} {\bf
  96} (2006) 032002 [\href{http://arXiv.org/abs/hep-ph/0510099}{{\tt
  hep-ph/0510099}}].
%%CITATION = HEP-PH/0510099;%%

\bibitem{Bjorken:1966jh}
J.~D. Bjorken, {\it {Applications of the Chiral U(6) x (6) Algebra of Current
  Densities}},  {\em Phys. Rev.} {\bf 148} (1966) 1467--1478.
%%CITATION = PHRVA,148,1467;%%

\bibitem{Bjorken:1969mm}
J.~D. Bjorken, {\it {Inelastic Scattering of Polarized Leptons from Polarized
  Nucleons}},  {\em Phys. Rev. D} {\bf 1} (1970) 1376--1379.

\bibitem{Larin:1990zw}
S.~Larin, F.~Tkachov and J.~Vermaseren, {\it {The alpha(s**3) correction to the
  Bjorken sum rule}},  {\em Phys. Rev. Lett.} {\bf 66} (1991) 862--863.

\bibitem{Tanabashi:2018oca}
{\bf Particle Data Group} Collaboration, M.~Tanabashi {\em et.~al.}, {\it
  {Review of Particle Physics}},  {\em Phys. Rev.} {\bf D98} (2018), no.~3
  030001.
%%CITATION = PHRVA,D98,030001;%%

\bibitem{Seng:2018yzq}
C.-Y. Seng, M.~Gorchtein, H.~H. Patel and M.~J. Ramsey-Musolf, {\it {Reduced
  Hadronic Uncertainty in the Determination of $V_{ud}$}},  {\em Phys. Rev.
  Lett.} {\bf 121} (2018), no.~24 241804
  [\href{http://arXiv.org/abs/1807.10197}{{\tt 1807.10197}}].
%%CITATION = ARXIV:1807.10197;%%

\bibitem{Seng:2018qru}
C.~Y. Seng, M.~Gorchtein and M.~J. Ramsey-Musolf, {\it {Dispersive evaluation
  of the inner radiative correction in neutron and nuclear $\beta$ decay}},
  {\em Phys. Rev.} {\bf D100} (2019), no.~1 013001
  [\href{http://arXiv.org/abs/1812.03352}{{\tt 1812.03352}}].
%%CITATION = ARXIV:1812.03352;%%

\bibitem{Nachtmann:1973mr}
O.~Nachtmann, {\it {Positivity constraints for anomalous dimensions}},  {\em
  Nucl. Phys. B} {\bf 63} (1973) 237--247.

\bibitem{Nachtmann:1974aj}
O.~Nachtmann, {\it {Is There Evidence for Large Anomalous Dimensions?}},  {\em
  Nucl. Phys. B} {\bf 78} (1974) 455--467.

\bibitem{Ye:2017gyb}
Z.~Ye, J.~Arrington, R.~J. Hill and G.~Lee, {\it {Proton and Neutron
  Electromagnetic Form Factors and Uncertainties}},  {\em Phys. Lett. B} {\bf
  777} (2018) 8--15 [\href{http://arXiv.org/abs/1707.09063}{{\tt 1707.09063}}].

\bibitem{Bhattacharya:2011ah}
B.~Bhattacharya, R.~J. Hill and G.~Paz, {\it {Model independent determination
  of the axial mass parameter in quasielastic neutrino-nucleon scattering}},
  {\em Phys. Rev. D} {\bf 84} (2011) 073006
  [\href{http://arXiv.org/abs/1108.0423}{{\tt 1108.0423}}].

\bibitem{Drechsel:2007if}
D.~Drechsel, S.~S. Kamalov and L.~Tiator, {\it {Unitary Isobar Model -
  MAID2007}},  {\em Eur. Phys. J. A} {\bf 34} (2007) 69--97
  [\href{http://arXiv.org/abs/0710.0306}{{\tt 0710.0306}}].

\bibitem{Lalakulich:2006sw}
O.~Lalakulich, E.~A. Paschos and G.~Piranishvili, {\it {Resonance production by
  neutrinos: The Second resonance region}},  {\em Phys. Rev. D} {\bf 74} (2006)
  014009 [\href{http://arXiv.org/abs/hep-ph/0602210}{{\tt hep-ph/0602210}}].

\bibitem{Kataev:1994rj}
A.~L. Kataev and A.~V. Sidorov, {\it {The Jacobi polynomials QCD analysis of
  the CCFR data for xF3 and the Q**2 dependence of the Gross-Llewellyn-Smith
  sum rule}},  {\em Phys. Lett.} {\bf B331} (1994) 179--186
  [\href{http://arXiv.org/abs/hep-ph/9402342}{{\tt hep-ph/9402342}}].
%%CITATION = HEP-PH/9402342;%%

\bibitem{Kim:1998kia}
J.~H. Kim {\em et.~al.}, {\it {A Measurement of alpha(s)(Q**2) from the
  Gross-Llewellyn Smith sum rule}},  {\em Phys. Rev. Lett.} {\bf 81} (1998)
  3595--3598 [\href{http://arXiv.org/abs/hep-ex/9808015}{{\tt
  hep-ex/9808015}}].
%%CITATION = HEP-EX/9808015;%%

\bibitem{Bolognese:1982zd}
{\bf Aachen-Bonn-CERN-Democritos-London-Oxford-Saclay} Collaboration,
  T.~Bolognese, P.~Fritze, J.~Morfin, D.~H. Perkins, K.~Powell and W.~G. Scott,
  {\it {Data on the Gross-llewellyn Smith Sum Rule as a Function of $q^2$}},
  {\em Phys. Rev. Lett.} {\bf 50} (1983) 224.
%%CITATION = PRLTA,50,224;%%

\bibitem{Allasia:1985hw}
D.~Allasia {\em et.~al.}, {\it {Q**2 Dependence of the Proton and Neutron
  Structure Functions from Neutrino and anti-neutrinos Scattering in
  Deuterium}},  {\em Z. Phys.} {\bf C28} (1985) 321.
%%CITATION = ZEPYA,C28,321;%%

\bibitem{deSwart:1963pdg}
J.~J. de~Swart, {\it {The Octet model and its Clebsch-Gordan coefficients}},
  {\em Rev. Mod. Phys.} {\bf 35} (1963) 916--939. [Erratum: Rev.Mod.Phys. 37,
  326--326 (1965)].

\bibitem{Czarnecki:2019mwq}
A.~Czarnecki, W.~J. Marciano and A.~Sirlin, {\it {Radiative Corrections to
  Neutron and Nuclear Beta Decays Revisited}},
  \href{http://arXiv.org/abs/1907.06737}{{\tt 1907.06737}}.
%%CITATION = ARXIV:1907.06737;%%

\bibitem{Seng:2020wjq}
C.-Y. Seng, X.~Feng, M.~Gorchtein and L.-C. Jin, {\it {Joint lattice
  QCD--dispersion theory analysis confirms the quark-mixing top-row unitarity
  deficit}},  {\em Phys. Rev. D} {\bf 101} (2020), no.~11 111301
  [\href{http://arXiv.org/abs/2003.11264}{{\tt 2003.11264}}].

\bibitem{Shiells:2020fqp}
K.~Shiells, P.~Blunden and W.~Melnitchouk, {\it {Electroweak axial structure
  functions and improved extraction of the $V_{ud}$ CKM matrix element}},
  \href{http://arXiv.org/abs/2012.01580}{{\tt 2012.01580}}.

\bibitem{Acciarri:2016crz}
{\bf DUNE} Collaboration, R.~Acciarri {\em et.~al.}, {\it {Long-Baseline
  Neutrino Facility (LBNF) and Deep Underground Neutrino Experiment (DUNE)}:
  {Conceptual Design Report, Volume 1: The LBNF and DUNE Projects}},
  \href{http://arXiv.org/abs/1601.05471}{{\tt 1601.05471}}.

\bibitem{Alvarez-Ruso:2017oui}
{\bf NuSTEC} Collaboration, L.~Alvarez-Ruso {\em et.~al.}, {\it {NuSTEC White
  Paper: Status and challenges of neutrino\textendash{}nucleus scattering}},
  {\em Prog. Part. Nucl. Phys.} {\bf 100} (2018) 1--68
  [\href{http://arXiv.org/abs/1706.03621}{{\tt 1706.03621}}].

\bibitem{Seng:2021qdx}
C.-Y. Seng, {\it {$V_{ud}$ radiative corrections with lattice input}},  in {\em
  {55th Rencontres de Moriond on Electroweak Interactions and Unified
  Theories}}, 4, 2021.
\newblock \href{http://arXiv.org/abs/2104.02586}{{\tt 2104.02586}}.

\bibitem{Baikov:2010je}
P.~Baikov, K.~Chetyrkin and J.~Kuhn, {\it {Adler Function, Bjorken Sum Rule,
  and the Crewther Relation to Order $\alpha^4_s$ in a General Gauge Theory}},
  {\em Phys. Rev. Lett.} {\bf 104} (2010) 132004
  [\href{http://arXiv.org/abs/1001.3606}{{\tt 1001.3606}}].

\bibitem{Baikov:2010iw}
P.~A. Baikov, K.~G. Chetyrkin and J.~H. Kuhn, {\it {Adler Function, DIS sum
  rules and Crewther Relations}},  {\em Nucl. Phys. B Proc. Suppl.} {\bf
  205-206} (2010) 237--241 [\href{http://arXiv.org/abs/1007.0478}{{\tt
  1007.0478}}].

\bibitem{Seng:2019plg}
C.-Y. Seng and U.-G. Mei\ss{}ner, {\it {Toward a First-Principles Calculation
  of Electroweak Box Diagrams}},  {\em Phys. Rev. Lett.} {\bf 122} (2019),
  no.~21 211802 [\href{http://arXiv.org/abs/1903.07969}{{\tt 1903.07969}}].
%%CITATION = ARXIV:1903.07969;%%

\bibitem{Sirlin:1967zza}
A.~Sirlin, {\it {General Properties of the Electromagnetic Corrections to the
  Beta Decay of a Physical Nucleon}},  {\em Phys. Rev.} {\bf 164} (1967)
  1767--1775.

\bibitem{Towner:2007np}
I.~S. Towner and J.~C. Hardy, {\it {An Improved calculation of the
  isospin-symmetry-breaking corrections to superallowed Fermi beta decay}},
  {\em Phys. Rev. C} {\bf 77} (2008) 025501
  [\href{http://arXiv.org/abs/0710.3181}{{\tt 0710.3181}}].

\bibitem{Sirlin:1987sy}
A.~Sirlin, {\it {Remarks Concerning the O(z alpha**2) Corrections to Fermi
  Decays, Conserved Vector Current Predictions and Universality}},  {\em Phys.
  Rev. D} {\bf 35} (1987) 3423.

\bibitem{Sirlin:1986cc}
A.~Sirlin and R.~Zucchini, {\it {Accurate Verification of the Conserved Vector
  Current and Standard Model Predictions}},  {\em Phys. Rev. Lett.} {\bf 57}
  (1986) 1994--1997.

\bibitem{Hardy:2014qxa}
J.~C. Hardy and I.~S. Towner, {\it {Superallowed $0^+\to 0^+$ nuclear
  \ensuremath{\beta} decays: 2014 critical survey, with precise results for
  $V_{ud}$ and CKM unitarity}},  {\em Phys. Rev. C} {\bf 91} (2015), no.~2
  025501 [\href{http://arXiv.org/abs/1411.5987}{{\tt 1411.5987}}].

\bibitem{Towner:1994mw}
I.~S. Towner, {\it {Quenching of spin operators in the calculation of radiative
  corrections for nuclear beta decay}},  {\em Phys. Lett. B} {\bf 333} (1994)
  13--16 [\href{http://arXiv.org/abs/nucl-th/9405031}{{\tt nucl-th/9405031}}].

\bibitem{Towner:2002rg}
I.~S. Towner and J.~C. Hardy, {\it {Calculated corrections to superallowed
  Fermi beta decay: New evaluation of the nuclear structure dependent terms}},
  {\em Phys. Rev. C} {\bf 66} (2002) 035501
  [\href{http://arXiv.org/abs/nucl-th/0209014}{{\tt nucl-th/0209014}}].

\bibitem{Gorchtein:2018fxl}
M.~Gorchtein, {\it {$\gamma W$ Box Inside Out: Nuclear Polarizabilities Distort
  the Beta Decay Spectrum}},  {\em Phys. Rev. Lett.} {\bf 123} (2019), no.~4
  042503 [\href{http://arXiv.org/abs/1812.04229}{{\tt 1812.04229}}].
%%CITATION = ARXIV:1812.04229;%%

\bibitem{Hardy:2008gy}
J.~C. Hardy and I.~S. Towner, {\it {Superallowed 0+ ---\ensuremath{>} 0+
  nuclear beta decays: A New survey with precision tests of the conserved
  vector current hypothesis and the standard model}},  {\em Phys. Rev. C} {\bf
  79} (2009) 055502 [\href{http://arXiv.org/abs/0812.1202}{{\tt 0812.1202}}].

\bibitem{Ormand:1989hm}
W.~E. Ormand and B.~A. Brown, {\it {Corrections to the Fermi Matrix Element for
  Superallowed Beta Decay}},  {\em Phys. Rev. Lett.} {\bf 62} (1989) 866--869.

\bibitem{Ormand:1995df}
W.~E. Ormand and B.~A. Brown, {\it {Isospin-mixing corrections for fp-shell
  Fermi transitions}},  {\em Phys. Rev. C} {\bf 52} (1995) 2455--2460
  [\href{http://arXiv.org/abs/nucl-th/9504017}{{\tt nucl-th/9504017}}].

\bibitem{Satula:2011br}
W.~Satula, J.~Dobaczewski, W.~Nazarewicz and M.~Rafalski, {\it {Microscopic
  calculations of isospin-breaking corrections to superallowed beta-decay}},
  {\em Phys. Rev. Lett.} {\bf 106} (2011) 132502
  [\href{http://arXiv.org/abs/1101.0939}{{\tt 1101.0939}}].

\bibitem{Liang:2009pf}
H.~Liang, N.~Van~Giai and J.~Meng, {\it {Isospin corrections for superallowed
  Fermi beta decay in self-consistent relativistic random-phase approximation
  approaches}},  {\em Phys. Rev. C} {\bf 79} (2009) 064316
  [\href{http://arXiv.org/abs/0904.3673}{{\tt 0904.3673}}].

\bibitem{Auerbach:2008ut}
N.~Auerbach, {\it {Coulomb corrections to superallowed beta decay in nuclei}},
  {\em Phys. Rev. C} {\bf 79} (2009) 035502
  [\href{http://arXiv.org/abs/0811.4742}{{\tt 0811.4742}}].

\bibitem{Damgaard:1969yyx}
J.~Damgaard, {\it {Corrections to the $ft$-values of $0^+ → 0^+$ superallowed
  \ensuremath{\beta}-decays}},  {\em Nucl. Phys. A} {\bf 130} (1969) 233--240.

\bibitem{Weinberg:1958ut}
S.~Weinberg, {\it {Charge symmetry of weak interactions}},  {\em Phys. Rev.}
  {\bf 112} (1958) 1375--1379.

\bibitem{Miller:2008my}
G.~A. Miller and A.~Schwenk, {\it {Isospin-symmetry-breaking corrections to
  superallowed Fermi beta decay: Formalism and schematic models}},  {\em Phys.
  Rev. C} {\bf 78} (2008) 035501 [\href{http://arXiv.org/abs/0805.0603}{{\tt
  0805.0603}}].

\bibitem{Miller:2009cg}
G.~A. Miller and A.~Schwenk, {\it {Isospin-symmetry-breaking corrections to
  superallowed Fermi beta decay: Radial excitations}},  {\em Phys. Rev. C} {\bf
  80} (2009) 064319 [\href{http://arXiv.org/abs/0910.2790}{{\tt 0910.2790}}].

\bibitem{Koshchii:2020qkr}
O.~Koshchii, J.~Erler, M.~Gorchtein, C.~J. Horowitz, J.~Piekarewicz,
  X.~Roca-Maza, C.-Y. Seng and H.~Spiesberger, {\it {Weak charge and weak
  radius of $^{12}$C}},  {\em Phys. Rev. C} {\bf 102} (2020), no.~2 022501
  [\href{http://arXiv.org/abs/2005.00479}{{\tt 2005.00479}}].

\bibitem{Marciano:2004uf}
W.~J. Marciano, {\it {Precise determination of |V(us)| from lattice
  calculations of pseudoscalar decay constants}},  {\em Phys. Rev. Lett.} {\bf
  93} (2004) 231803 [\href{http://arXiv.org/abs/hep-ph/0402299}{{\tt
  hep-ph/0402299}}].

\bibitem{Cirigliano:2011tm}
V.~Cirigliano and H.~Neufeld, {\it {A note on isospin violation in Pl2(gamma)
  decays}},  {\em Phys. Lett. B} {\bf 700} (2011) 7--10
  [\href{http://arXiv.org/abs/1102.0563}{{\tt 1102.0563}}].

\bibitem{Aoki:2021kgd}
Y.~Aoki {\em et.~al.}, {\it {FLAG Review 2021}},
  \href{http://arXiv.org/abs/2111.09849}{{\tt 2111.09849}}.

\bibitem{Bazavov:2017lyh}
A.~Bazavov {\em et.~al.}, {\it {$B$- and $D$-meson leptonic decay constants
  from four-flavor lattice QCD}},  {\em Phys. Rev. D} {\bf 98} (2018), no.~7
  074512 [\href{http://arXiv.org/abs/1712.09262}{{\tt 1712.09262}}].

\bibitem{Dowdall:2013rya}
R.~J. Dowdall, C.~T.~H. Davies, G.~P. Lepage and C.~McNeile, {\it {Vus from pi
  and K decay constants in full lattice QCD with physical u, d, s and c
  quarks}},  {\em Phys. Rev. D} {\bf 88} (2013) 074504
  [\href{http://arXiv.org/abs/1303.1670}{{\tt 1303.1670}}].

\bibitem{Carrasco:2014poa}
N.~Carrasco {\em et.~al.}, {\it {Leptonic decay constants $f_{K},f_{D},$ and
  $f_{{D}_{s}}$ with $N_{f} = 2+1+1$ twisted-mass lattice QCD}},  {\em Phys.
  Rev. D} {\bf 91} (2015), no.~5 054507
  [\href{http://arXiv.org/abs/1411.7908}{{\tt 1411.7908}}].

\bibitem{Miller:2020xhy}
N.~Miller {\em et.~al.}, {\it {$F_K / F_\pi$ from M\"obius Domain-Wall fermions
  solved on gradient-flowed HISQ ensembles}},  {\em Phys. Rev. D} {\bf 102}
  (2020), no.~3 034507 [\href{http://arXiv.org/abs/2005.04795}{{\tt
  2005.04795}}].

\bibitem{Giusti:2017dwk}
D.~Giusti, V.~Lubicz, G.~Martinelli, C.~T. Sachrajda, F.~Sanfilippo, S.~Simula,
  N.~Tantalo and C.~Tarantino, {\it {First lattice calculation of the QED
  corrections to leptonic decay rates}},  {\em Phys. Rev. Lett.} {\bf 120}
  (2018), no.~7 072001 [\href{http://arXiv.org/abs/1711.06537}{{\tt
  1711.06537}}].
%%CITATION = ARXIV:1711.06537;%%

\bibitem{DiCarlo:2019thl}
M.~Di~Carlo, D.~Giusti, V.~Lubicz, G.~Martinelli, C.~T. Sachrajda,
  F.~Sanfilippo, S.~Simula and N.~Tantalo, {\it {Light-meson leptonic decay
  rates in lattice QCD+QED}},  {\em Phys. Rev.} {\bf D100} (2019), no.~3 034514
  [\href{http://arXiv.org/abs/1904.08731}{{\tt 1904.08731}}].
%%CITATION = ARXIV:1904.08731;%%

\bibitem{KLOE:2005vdt}
{\bf KLOE} Collaboration, F.~Ambrosino {\em et.~al.}, {\it {Measurements of the
  absolute branching ratios for the dominant K(L) decays, the K(L) lifetime,
  and V(us) with the KLOE detector}},  {\em Phys. Lett. B} {\bf 632} (2006)
  43--50 [\href{http://arXiv.org/abs/hep-ex/0508027}{{\tt hep-ex/0508027}}].

\bibitem{KTeV:2004hpx}
{\bf KTeV} Collaboration, T.~Alexopoulos {\em et.~al.}, {\it {Measurements of
  K(L) branching fractions and the CP violation parameter |eta+-|}},  {\em
  Phys. Rev. D} {\bf 70} (2004) 092006
  [\href{http://arXiv.org/abs/hep-ex/0406002}{{\tt hep-ex/0406002}}].

\bibitem{Batley:2007zzb}
J.~R. Batley {\em et.~al.}, {\it {Determination of the relative decay rate K(S)
  ---\ensuremath{>} pi e nu / K(L) ---\ensuremath{>} pi e nu}},  {\em Phys.
  Lett. B} {\bf 653} (2007) 145--150.

\bibitem{KLOE:2006vvm}
{\bf KLOE} Collaboration, F.~Ambrosino {\em et.~al.}, {\it {Study of the
  branching ratio and charge asymmetry for the decay $K(s) \to \pi e \nu$ with
  the KLOE detector}},  {\em Phys. Lett. B} {\bf 636} (2006) 173--182
  [\href{http://arXiv.org/abs/hep-ex/0601026}{{\tt hep-ex/0601026}}].

\bibitem{KLOE:2002lao}
{\bf KLOE} Collaboration, A.~Aloisio {\em et.~al.}, {\it {Measurement of the
  branching fraction for the decay $K(S) \to \pi e \nu$}},  {\em Phys. Lett. B}
  {\bf 535} (2002) 37--42 [\href{http://arXiv.org/abs/hep-ph/0203232}{{\tt
  hep-ph/0203232}}].

\bibitem{KLOE:2007jte}
{\bf KLOE} Collaboration, F.~Ambrosino {\em et.~al.}, {\it {Measurement of the
  absolute branching ratios for semileptonic $K^\pm$ decays with the KLOE
  detector}},  {\em JHEP} {\bf 02} (2008) 098
  [\href{http://arXiv.org/abs/0712.3841}{{\tt 0712.3841}}].

\bibitem{Chiang:1972rp}
I.~H. Chiang, J.~L. Rosen, S.~Shapiro, R.~Handler, S.~Olsen and L.~Pondrom,
  {\it {$K^+$ Decay in Flight}},  {\em Phys. Rev. D} {\bf 6} (1972) 1254.

\bibitem{Marciano:1993sh}
W.~J. Marciano and A.~Sirlin, {\it {Radiative corrections to pi(lepton 2)
  decays}},  {\em Phys. Rev. Lett.} {\bf 71} (1993) 3629--3632.

\bibitem{Cirigliano:2008wn}
V.~Cirigliano, M.~Giannotti and H.~Neufeld, {\it {Electromagnetic effects in
  K(l3) decays}},  {\em JHEP} {\bf 11} (2008) 006
  [\href{http://arXiv.org/abs/0807.4507}{{\tt 0807.4507}}].
%%CITATION = ARXIV:0807.4507;%%

\bibitem{Cirigliano:2011ny}
V.~Cirigliano, G.~Ecker, H.~Neufeld, A.~Pich and J.~Portoles, {\it {Kaon Decays
  in the Standard Model}},  {\em Rev. Mod. Phys.} {\bf 84} (2012) 399
  [\href{http://arXiv.org/abs/1107.6001}{{\tt 1107.6001}}].
%%CITATION = ARXIV:1107.6001;%%

\bibitem{Cirigliano:2001mk}
V.~Cirigliano, M.~Knecht, H.~Neufeld, H.~Rupertsberger and P.~Talavera, {\it
  {Radiative corrections to K(l3) decays}},  {\em Eur. Phys. J.} {\bf C23}
  (2002) 121--133 [\href{http://arXiv.org/abs/hep-ph/0110153}{{\tt
  hep-ph/0110153}}].
%%CITATION = HEP-PH/0110153;%%

\bibitem{Cirigliano:2004pv}
V.~Cirigliano, H.~Neufeld and H.~Pichl, {\it {K(e3) decays and CKM unitarity}},
   {\em Eur. Phys. J. C} {\bf 35} (2004) 53--65
  [\href{http://arXiv.org/abs/hep-ph/0401173}{{\tt hep-ph/0401173}}].

\bibitem{Antonelli:2010yf}
{\bf FlaviaNet Working Group on Kaon Decays} Collaboration, M.~Antonelli {\em
  et.~al.}, {\it {An Evaluation of $|V_{us}|$ and precise tests of the Standard
  Model from world data on leptonic and semileptonic kaon decays}},  {\em Eur.
  Phys. J. C} {\bf 69} (2010) 399--424
  [\href{http://arXiv.org/abs/1005.2323}{{\tt 1005.2323}}].

\bibitem{Carrasco:2016kpy}
N.~Carrasco, P.~Lami, V.~Lubicz, L.~Riggio, S.~Simula and C.~Tarantino, {\it
  {$K \to \pi$ semileptonic form factors with $N_f=2+1+1$ twisted mass
  fermions}},  {\em Phys. Rev. D} {\bf 93} (2016), no.~11 114512
  [\href{http://arXiv.org/abs/1602.04113}{{\tt 1602.04113}}].

\bibitem{Bazavov:2018kjg}
{\bf Fermilab Lattice, MILC} Collaboration, A.~Bazavov {\em et.~al.}, {\it
  {$|V_{us}|$ from $K_{\ell 3}$ decay and four-flavor lattice QCD}},  {\em
  Phys. Rev.} {\bf D99} (2019), no.~11 114509
  [\href{http://arXiv.org/abs/1809.02827}{{\tt 1809.02827}}].
%%CITATION = ARXIV:1809.02827;%%

\bibitem{Kakazu:2019ltq}
{\bf PACS} Collaboration, J.~Kakazu, K.-i. Ishikawa, N.~Ishizuka, Y.~Kuramashi,
  Y.~Nakamura, Y.~Namekawa, Y.~Taniguchi, N.~Ukita, T.~Yamazaki and
  T.~Yoshi\'e, {\it {$K_{l3}$ form factors at the physical point on a $(10.9
  fm)^3$ volume}},  {\em Phys. Rev. D} {\bf 101} (2020), no.~9 094504
  [\href{http://arXiv.org/abs/1912.13127}{{\tt 1912.13127}}].

\bibitem{Yamazaki:2021zxz}
{\bf PACS} Collaboration, T.~Yamazaki, K.-i. Ishikawa, N.~Ishizuka,
  Y.~Kuramashi, Y.~Nakamura, Y.~Namekawa, Y.~Taniguchi, N.~Ukita and
  T.~Yoshi\'e, {\it {Calculation of kaon semileptonic form factor with the
  PACS10 configuration}},  in {\em {38th International Symposium on Lattice
  Field Theory}}, 11, 2021.
\newblock \href{http://arXiv.org/abs/2111.00744}{{\tt 2111.00744}}.

\bibitem{PACStalk}
T.~Yamazaki, 2021.
\newblock Calculation of kaon semileptonic form factor with the PACS10
  configurations, \textit{The 38th International Symposium on Lattice Field
  Theory (LATTICE21)},
  \url{https://indico.cern.ch/event/1006302/contributions/4373348/attachments/2287737/3888678/yamazaki.pdf}.

\bibitem{Hill:2006bq}
R.~J. Hill, {\it {Constraints on the form factors for K ---\ensuremath{>} pi l
  nu and implications for |V(us)|}},  {\em Phys. Rev. D} {\bf 74} (2006) 096006
  [\href{http://arXiv.org/abs/hep-ph/0607108}{{\tt hep-ph/0607108}}].

\bibitem{Lichard:1997ya}
P.~Lichard, {\it {Some implications of meson dominance in weak interactions}},
  {\em Phys. Rev. D} {\bf 55} (1997) 5385--5407
  [\href{http://arXiv.org/abs/hep-ph/9702345}{{\tt hep-ph/9702345}}].

\bibitem{Bernard:2006gy}
V.~Bernard, M.~Oertel, E.~Passemar and J.~Stern, {\it {K(mu3)**L decay: A
  Stringent test of right-handed quark currents}},  {\em Phys. Lett.} {\bf
  B638} (2006) 480--486 [\href{http://arXiv.org/abs/hep-ph/0603202}{{\tt
  hep-ph/0603202}}].
%%CITATION = HEP-PH/0603202;%%

\bibitem{Bernard:2009zm}
V.~Bernard, M.~Oertel, E.~Passemar and J.~Stern, {\it {Dispersive
  representation and shape of the K(l3) form factors: Robustness}},  {\em Phys.
  Rev.} {\bf D80} (2009) 034034 [\href{http://arXiv.org/abs/0903.1654}{{\tt
  0903.1654}}].
%%CITATION = ARXIV:0903.1654;%%

\bibitem{Abouzaid:2009ry}
{\bf KTeV} Collaboration, E.~Abouzaid {\em et.~al.}, {\it {Dispersive analysis
  of K (L mu3) and K (L e3) scalar and vector form factors using KTeV data}},
  {\em Phys. Rev. D} {\bf 81} (2010) 052001
  [\href{http://arXiv.org/abs/0912.1291}{{\tt 0912.1291}}].

\bibitem{MoulsonCKM21}
M.~Moulson, 2021.
\newblock $V_{us}$ from kaon decays, \textit{11th International Workshop on the
  CKM Unitarity Triangle (CKM 2021)},
  \url{https://indico.cern.ch/event/891123/contributions/4601856/attachments/2351074/4011941/CKM%202021.pdf}.

\bibitem{Ananthanarayan:2004qk}
B.~Ananthanarayan and B.~Moussallam, {\it {Four-point correlator constraints on
  electromagnetic chiral parameters and resonance effective Lagrangians}},
  {\em JHEP} {\bf 06} (2004) 047
  [\href{http://arXiv.org/abs/hep-ph/0405206}{{\tt hep-ph/0405206}}].
%%CITATION = HEP-PH/0405206;%%

\bibitem{DescotesGenon:2005pw}
S.~Descotes-Genon and B.~Moussallam, {\it {Radiative corrections in weak
  semi-leptonic processes at low energy: A Two-step matching determination}},
  {\em Eur. Phys. J.} {\bf C42} (2005) 403--417
  [\href{http://arXiv.org/abs/hep-ph/0505077}{{\tt hep-ph/0505077}}].
%%CITATION = HEP-PH/0505077;%%

\bibitem{Seng:2019lxf}
C.-Y. Seng, D.~Galviz and U.-G. Mei\ss{}ner, {\it {A New Theory Framework for
  the Electroweak Radiative Corrections in $K_{l3}$ Decays}},  {\em JHEP} {\bf
  02} (2020) 069 [\href{http://arXiv.org/abs/1910.13208}{{\tt 1910.13208}}].

\bibitem{Seng:2020jtz}
C.-Y. Seng, X.~Feng, M.~Gorchtein, L.-C. Jin and U.-G. Mei\ss{}ner, {\it {New
  method for calculating electromagnetic effects in semileptonic beta-decays of
  mesons}},  {\em JHEP} {\bf 10} (2020) 179
  [\href{http://arXiv.org/abs/2009.00459}{{\tt 2009.00459}}].

\bibitem{Ma:2021azh}
P.-X. Ma, X.~Feng, M.~Gorchtein, L.-C. Jin and C.-Y. Seng, {\it {Lattice QCD
  calculation of the electroweak box diagrams for the kaon semileptonic
  decays}},  {\em Phys. Rev. D} {\bf 103} (2021) 114503
  [\href{http://arXiv.org/abs/2102.12048}{{\tt 2102.12048}}].

\bibitem{Seng:2021boy}
C.-Y. Seng, D.~Galviz, M.~Gorchtein and U.~G. Mei\ss{}ner, {\it {High-precision
  determination of the Ke3 radiative corrections}},  {\em Phys. Lett. B} {\bf
  820} (2021) 136522 [\href{http://arXiv.org/abs/2103.00975}{{\tt
  2103.00975}}].

\bibitem{Seng:2021wcf}
C.-Y. Seng, D.~Galviz, M.~Gorchtein and U.-G. Mei\ss{}ner, {\it {Improved
  $K_{e3}$ radiative corrections sharpen the $K_{\mu 2}$\textendash{}K$_{l3}$
  discrepancy}},  {\em JHEP} {\bf 11} (2021) 172
  [\href{http://arXiv.org/abs/2103.04843}{{\tt 2103.04843}}].

\bibitem{RBC:2014ntl}
{\bf RBC, UKQCD} Collaboration, T.~Blum {\em et.~al.}, {\it {Domain wall QCD
  with physical quark masses}},  {\em Phys. Rev. D} {\bf 93} (2016), no.~7
  074505 [\href{http://arXiv.org/abs/1411.7017}{{\tt 1411.7017}}].

\bibitem{Durr:2010vn}
S.~Durr, Z.~Fodor, C.~Hoelbling, S.~D. Katz, S.~Krieg, T.~Kurth, L.~Lellouch,
  T.~Lippert, K.~K. Szabo and G.~Vulvert, {\it {Lattice QCD at the physical
  point: light quark masses}},  {\em Phys. Lett. B} {\bf 701} (2011) 265--268
  [\href{http://arXiv.org/abs/1011.2403}{{\tt 1011.2403}}].

\bibitem{Durr:2010aw}
S.~Durr, Z.~Fodor, C.~Hoelbling, S.~D. Katz, S.~Krieg, T.~Kurth, L.~Lellouch,
  T.~Lippert, K.~K. Szabo and G.~Vulvert, {\it {Lattice QCD at the physical
  point: Simulation and analysis details}},  {\em JHEP} {\bf 08} (2011) 148
  [\href{http://arXiv.org/abs/1011.2711}{{\tt 1011.2711}}].

\bibitem{MILC:2009ltw}
{\bf MILC} Collaboration, A.~Bazavov {\em et.~al.}, {\it {MILC results for
  light pseudoscalars}},  {\em PoS} {\bf CD09} (2009) 007
  [\href{http://arXiv.org/abs/0910.2966}{{\tt 0910.2966}}].

\bibitem{Fodor:2016bgu}
Z.~Fodor, C.~Hoelbling, S.~Krieg, L.~Lellouch, T.~Lippert, A.~Portelli,
  A.~Sastre, K.~K. Szabo and L.~Varnhorst, {\it {Up and down quark masses and
  corrections to Dashen's theorem from lattice QCD and quenched QED}},  {\em
  Phys. Rev. Lett.} {\bf 117} (2016), no.~8 082001
  [\href{http://arXiv.org/abs/1604.07112}{{\tt 1604.07112}}].

\bibitem{Colangelo:2018jxw}
G.~Colangelo, S.~Lanz, H.~Leutwyler and E.~Passemar, {\it {Dispersive analysis
  of $\eta \rightarrow 3 \pi $}},  {\em Eur. Phys. J. C} {\bf 78} (2018),
  no.~11 947 [\href{http://arXiv.org/abs/1807.11937}{{\tt 1807.11937}}].

\end{thebibliography}\endgroup

\end{document}